\documentclass[acmsmall, nonacm]{acmart}
\setcopyright{none}

\renewcommand\footnotetextcopyrightpermission[1]{}
\usepackage{filecontents}
\usepackage{color}
\usepackage{xspace}
\usepackage{comment}
\usepackage{colortbl}
\usepackage{graphicx}
\usepackage{float}
\usepackage{multirow}
\usepackage{makecell}
\usepackage[ruled,vlined]{algorithm2e}

\newcommand{\tool}{\textsc{Boian}\xspace}
\newcommand{\ossfuzz}{OSS-Fuzz\xspace}

\newcommand{\benchmark}{\textsc{UbuntuBench}\xspace}

\newcommand{\custombox}[2]{
\vspace{5pt}
\begin{center}
\fbox{\parbox{0.95\columnwidth}{\textit{Result {#1}: {#2}}}}
\end{center}
\vspace{10pt}
}

\newcommand{\comm}[1]{\hfill \textcolor{blue}{\(\triangleright\) \textit{#1}}}

\newcommand{\comz}[1]{ \textcolor{blue}{\(\triangleright\) \textit{#1}}}

\newcommand{\tprogs}{675\xspace}
\newcommand{\tbugs}{4,908\xspace}

\newcommand{\code}[1]{\texttt{#1}}

\begin{document}

\date{}

\title{Fuzzing at Scale: The Untold Story of the Scheduler}

\author{Ivica Nikoli{\'c}}
\affiliation{%
  \institution{National University of Singapore}
  \country{Singapore}}
\email{inikolic@nus.edu.sg}

\author{Racchit Jain}
\authornote{Work done while interning at National University of Singapore.}
\affiliation{%
  \institution{National University of Singapore}
  \country{Singapore}}
\email{racchit.jain@gmail.com}

\begin{abstract}
\emph{How to search for bugs in 1,000 programs using a pre-existing fuzzer and a standard PC?} 
We consider this problem and show that a well-designed strategy that determines which programs to fuzz and for how long can greatly impact the number of bugs found across the programs.
In fact, the impact of employing an effective strategy is comparable to that of utilizing a state-of-the-art fuzzer.
The considered problem is referred to as  \emph{fuzzing at scale}, and the strategy as \emph{scheduler}. 
We show that besides a naive scheduler,  that allocates equal fuzz time to all programs, 
we can consider dynamic schedulers that adjust time allocation based on the ongoing fuzzing progress of individual programs. Such schedulers are superior because they lead both to higher number of total found bugs and to higher number of found bugs for most programs. 
The performance gap between naive and dynamic schedulers can be as wide (or even wider) as the gap between two fuzzers.
Our findings thus suggest that the problem of advancing schedulers is fundamental for fuzzing at scale. 
We develop several schedulers and leverage the most sophisticated one to fuzz simultaneously our newly compiled benchmark of around 5,000 Ubuntu programs, and detect \tbugs bugs.
\end{abstract}

\maketitle

\section{Introduction}
\label{sect:intro}

Fuzzing is an effective software testing technique used for uncovering vulnerabilities. As software systems become increasingly complex and pervasive, ensuring their security and reliability is important. Despite the use of various testing methodologies, software bugs and vulnerabilities continue to emerge, posing significant risks to both organizations and individuals relying on the software. Consequently, the demand for more efficient and effective testing techniques has grown, with fuzzing emerging as a leading approach to address these concerns. Fuzzing can identify potential security flaws and weaknesses in software applications by providing unexpected, invalid, or random inputs (called testcases) to the target software, potentially revealing issues that might have remained hidden under normal inputs.

The primary focus of research in fuzzing has been on the development and enhancement of fuzzers, as evidenced by numerous designs and  studies~\cite{zalewski2017american,fioraldi2020afl,swiecki2021honggfuzz,koike2022slopt,nikolic2021refined,gan2018collafl,aschermann2019redqueen,osterlund2020parmesan,yun2018qsym,rawat2017vuzzer,chen2018angora,liang2022pata,lyu2019mopt,lyu2022ems,liang2019deepfuzzer,lemieux2018fairfuzz,chen2019enfuzz,yue2020ecofuzz,she2022effective,geretto2022snappy,yu2022htfuzz,lee2021constraint,chen2020muzz,gan2020greyone,zong2020fuzzguard,nguyen2020binary,chen2020meuzz,nagy2021breaking,zhou2022ferry,wangcarpetfuzz,blazytko2019grimoire,shah2022mc2,zhu2021regression,nagy2021same,chen2019matryoshka,cho2019intriguer}. This emphasis is well-aligned with the principal use case of fuzzing, which typically involves testing a single program at a time. 
For example, a software developer writes a program and then employs a fuzzer to identify potential bugs in its code. The efficiency of a fuzzer, often assessed using metrics such as \emph{code coverage} (which measures the proportion of program code executed during fuzzing), directly impacts the likelihood of detecting potential bugs in the target program. Consequently, there has been a sustained effort to design increasingly efficient fuzzers. Each new fuzzer or fuzzing technique  demonstrates certain advantages over its predecessors. 
However, despite the constant progress, the disparity between the most effective fuzzers and their less efficient counterparts remains relatively small. For instance, FuzzBench, a platform developed by Google for comparing fuzzers~\cite{metzman2021fuzzbench}, reveals an average performance gap of only 20\%-30\% between the best and worst-performing fuzzers across multiple programs.

Our research objective deviates from the conventional emphasis on fuzzers. Instead, we pursue an orthogonal approach, focusing on developing techniques to utilize existing fuzzers for simultaneously testing multiple programs. We term this framework \emph{fuzzing at scale}. More specifically, given a large set of target programs and a fuzzer, 
our aim is to develop a \emph{scheduler} that will control the simultaneous fuzzing of all these programs across multiple CPU cores.
The goal of the scheduler is to facilitate \emph{efficient} fuzzing in an aggregated manner, that is, collectively across all the considered programs. For example, one objective could be to maximize the cumulative code coverage, which refers to the total sum of coverage achieved by the fuzzer for each program.
To the best of our knowledge, this fuzzing at scale paradigm with such an explicit objective has not been previously considered. 
In contrast, platforms for distributed fuzzing of multiple programs, such as Google's OSS-Fuzz~\cite{serebryany2017oss}, allocate computational resources (e.g., a designated fuzz time budget) equally among the fuzzed programs. Furthermore, this trivial approach, known as the \emph{baseline scheduler}, has been applied to systems where the computational resources per fuzzed program are fairly large, for instance, when one or several CPUs can be allocated to fuzz test a single program. However, the inverse scenario, in which the number of tested programs  surpasses the number of available CPU cores, has not been investigated.
Nonetheless, use cases of this type are not unusual. For instance, an organization like a defense company or government security agency might prefer first to fuzz all required Linux applications (which could number in the hundreds or even thousands) to address apparent bugs before utilizing this open-source software.  
A security researcher compiling a database of bugs across multitude of programs would benefit from fuzzing at scale by testing large number of programs at once. 
In fact, any fuzzing scenario where the number of target programs exceeds that of available CPUs is indeed fuzzing at scale -- for instance, fuzz testing 10 programs on 2 CPUs.

In this paper, we consider the fuzzing at scale framework and demonstrate the significance of scheduler selection. Our primary contribution reveals that allocating equal fuzz time to all programs, as the baseline scheduler does, is suboptimal. More advanced schedulers provides up to a 30\% advantage over the baseline, emphasizing that \emph{in fuzzing at scale, the choice of a scheduler is as critical as the choice of a fuzzer}. 
Furthermore, the combination of an outstanding scheduler with an average fuzzer may match or even surpass the performance of a baseline scheduler and an outstanding fuzzer. This observation highlights the need for research into effective schedulers. As an initial step in this direction, we develop several new schedulers, each grounded in the fundamental and widely-used fuzzing hypothesis that a program's past code coverage correlates to its future coverage. Consequently, these schedulers are designed to prioritize fuzzing programs that have previously resulted in high coverage, scheduling them for fuzzing more frequently.
Our final design, referred to as \tool, outperforms all other schedulers according to metrics such as total code coverage achieved across all programs and the percentage of programs for which the highest coverage has been attained.

Our research presents an initial exploration into the domain of fuzzing at scale. In order to evaluate and compare the efficiency of algorithms in this field, a comprehensive benchmark set is necessary. 
We assemble an extensive fuzzing benchmark, referred to as \benchmark, which consists of 5,467 programs derived from a collection of approximately 75,799 Ubuntu packages. This benchmark not only serves as the basis for our investigation, but also offers potential benefits for future research efforts in the area of fuzzing at scale. We employ this as well as the smaller \ossfuzz~\cite{ossfuzzbugs} benchmark
to assess schedulers and to demonstrate the applicability of our fuzzing at scale framework. More specifically, we utilize the newly developed \tool to schedule simultaneous fuzzing of all programs in \benchmark using state-of-the-art fuzzers AFL++~\cite{fioraldi2020afl} and Honggfuzz\cite{swiecki2021honggfuzz}. As a result, we identify \tbugs bugs among \tprogs programs within a three-day fuzzing period with each fuzzer on a 30-core machine.

\section{Problem}

Fuzzing at scale  consists of distributing fuzzing
effort across multiple cores and machines. 
It can be categorized into two distinct types: the first aims to uncover deep bugs, while the second aims to fuzz a wide range of programs.
For simplicity, we can refer to first type as \emph{deep}, while to the second as \emph{wide}. While prior fuzzing at scale work has been focused on the deep type, our objective in this study is to concentrate on the wide type.

Deep
fuzzing at scale uses potentially large computational  resources to fuzz test one program on bugs. The aim of such type of thorough fuzzing is to discover potential vulnerabilities hidden deeply in the program code, that usually require substantial time for detection. 
The primary use case of this rigorous fuzzing approach is to ensure the security of a small set of critical software components. These components often include the most popular binaries and libraries. The goal is to keep them free of vulnerabilities that are easy or moderately challenging to discover.

Wide 
fuzzing at scale uses the available resources to fuzz simultaneously a larger number of programs. In contrast to deep, 
the aim of wide fuzzing 
is to cover a wide array of target programs, ensuring that none possess easily discoverable vulnerabilities.  
\textbf{(But given resources, it can also uncover deeply hidden vulnerabilities, because it uses the same fuzzer as deep fuzzing.)}
When faced with a large set of programs, such as a Linux distribution containing numerous binaries and libraries of varying importance, it is challenging to decide which of the following two strategies is more vital: %
deep 
fuzzing of a small set of programs or 
wide 
fuzzing of all programs. The former strategy effectively hardens a small yet crucial set of programs but leaves even trivial bugs in the remaining programs. In contrast, the latter strategy may identify trivial bugs across all programs but may leave deeper bugs undiscovered \textbf{(assuming time constrained fuzzing, but longer wide fuzzing campaigns will detect them)}.
Given these considerations, it is reasonable to assume that both types of fuzzing at scale  are relevant.
However, in general the unit cost of bug detection through %
wide 
fuzzing at scale is significantly lower than that of 
deep
fuzzing because it is easier to discover some bug in large set of programs than a deep bug in small set. This is attributed to the diminishing returns of fuzzing; as a program undergoes more fuzz testing, fewer bugs are detected, and the cost of discovering each bug increases substantially. So, due to the larger number of targets, the 
wide 
approach will produce bugs faster than the 
deep. 
Hence, 
the wide type
is the more potent source of  vulnerabilities among the two fuzzing at scale types. 
Interestingly, only %
deep 
fuzzing at scale has been explored to date. Our objective is to shift the focus to the second and unexplored type, i.e., to  
wide fuzzing at scale. 

The effectiveness of 
deep
fuzzing at scale relies on the quality of the utilized fuzzer because this approach fuzz tests individual target programs one after another for an extended duration on a multi-core system. Since %
deep fuzzing 
operates on a single target at a time, improving its efficiency—in terms of a metric like code coverage—requires employing a superior fuzzer.
Conversely, 
wide fuzzing 
operates on multiple targets concurrently, meaning that the choice of fuzzer may not be the sole determinant of its efficiency. An additional factor to consider is the distribution of computational resources, particularly CPU time, among the fuzzed programs. A basic method involves allocating equal amounts of time to each program in a round-robin manner. For instance, fuzzing 40 programs on a 4-core machine for 10 hours would allocate one hour to each program.
However, equal allocation of time may not be the most efficient strategy, as it does not account for the inherent differences between programs that can result in uneven progress in code coverage across the target set. The diversity of programs can directly impact the rate at which a fuzzer explores and discovers new coverage. 
Thus, allocating an equal amount of time to all fuzzed programs, may lead to inefficient fuzzing, e.g., wasting resources on programs that have already been well examined, instead of on those that have shown consistent increase of code coverage. 
Inefficiency in fuzzing poses a particularly severe problem due to its inherent time-consuming nature, i.e. often fuzzing runs for days or even weeks, thereby amplifying the significance of resource wastage. 

To address this issue, a more adaptive and dynamic approach to time allocation should be considered, wherein fuzzing time is apportioned based on some feedback obtained from the previous fuzzing runs of target programs. More precisely, by monitoring the progress of code coverage and adjusting the fuzzing time accordingly, one can ensure a more efficient and comprehensive 
wide 
fuzzing at scale.

\vspace{10pt}
\noindent\textbf{Concrete applications.} 
To further motivate our focus on wide fuzzing at scale, let us consider several concrete applications:
\begin{itemize}
\item \textit{Linux distribution hardening:} Consider a new Linux distribution release, encompassing thousands of software packages.  A crucial task for the maintainers is to quickly identify and fix vulnerabilities across all these packages. Wide fuzzing at scale allows security researchers to perform a preliminary security assessment by running a short fuzzing campaign. This can help identify easily discoverable vulnerabilities early on, allowing for timely patching before the official release. In contrast, deep fuzzing at scale on a selected subset of critical packages might miss vulnerabilities in less critical ones, leaving the overall system exposed. 
\item \textit{Fuzzer comparison}:
When evaluating the performance of different fuzzers, wide fuzzing at scale may provide a more robust and diverse test environment than deep fuzzing on a limited set of programs. By fuzzing a large number of varied programs, researchers can obtain a more accurate assessment of a fuzzer's strengths and weaknesses across different program types and complexities. This approach allows for better differentiation between fuzzers and more reliable performance metrics, as it accounts for the fuzzer's adaptability to diverse codebases.
Testing a wider range of programs helps prevent potential overfitting, where a fuzzer might perform exceptionally well on a small set of specific programs but fail to generalize to other codebases. By evaluating fuzzers across many more programs, researchers can better measure their true performance and adaptability, reducing the risk of drawing conclusions based on limited sample set.
\item \textit{Vulnerability database compilation:} 
Using the best available fuzzers, and applied to a large number of programs, wide fuzzing can efficiently discover bugs across diverse software. This approach is particularly effective for building vulnerability databases because its primary objective is to maximize the number of bugs found. As a result, wide fuzzing can quickly build a broad, representative collection of vulnerabilities, creating a valuable resource for security research and software development. The resulting database can offer a comprehensive view of vulnerabilities across different software.
\item \textit{Fuzzing different program arguments:}
Wide fuzzing can be employed to fuzz test a program with different command line arguments. In this scenario, each fuzzer within the framework will be assigned to a specific set of fixed command line arguments. Wide fuzzing will utilize coverage feedback to guide the resource allocation across the different fuzzers -- it will favour and fuzz more the arguments that show progress in code coverage.
In contrast, a simple naive approach to fuzzing different command line arguments lacks this intelligent resource allocation mechanism. Without coverage feedback, it is not clear how to optimally distribute fuzzing efforts among various sets of fixed arguments, potentially leading to less efficient exploration of the program's behavior under different input arguments.
\item \textit{Security audit of large software repositories:} Consider an organization like a bank that utilizes hundreds of open-source libraries and tools within their software ecosystem. %
Before implementation, it is crucial to conduct a  security audit to identify and mitigate potential vulnerabilities in the libraries. 
Wide fuzzing at scale allows the organization to efficiently uncover superficial bugs across a wide range of programs within a shorter time frame. This approach ensures that all programs meet a minimum security baseline before deployment. 

\item \textit{Building fuzzing benchmarks and datasets:} Creating comprehensive fuzzing benchmarks requires collecting diverse programs and datasets, often involving hundreds or thousands of applications. Wide fuzzing at scale can help to automatically assess the difficulty of finding bugs in these programs. 
For instance, when building a new fuzzing benchmark, researchers can use wide fuzzing to quickly test a large pool of potential programs, analyzing the rate of bugs found per unit of time. This analysis would help in categorizing programs based on their resilience to fuzzing, enabling the creation of a benchmark with varying difficulty levels. 
\end{itemize}

\section{Fuzzing at Scale Framework}
\label{sect:approach}

In the remaining of the paper, we use \emph{fuzzing at scale} to refer exclusively to the 
wide  type of fuzzing at scale. 
Further, first we provide a definition of the fuzzing at scale problem under consideration. Subsequently, we introduce a variety of schedulers designed to address this problem. Finally, we provide additional details about particular components of the framework.

\subsection{Problem Setup and Objective}

Fuzzing at scale consists of  utilizing available computational resources to fuzz a  pool of programs. We are interested in improving the efficiency of this task.
More precisely, given $p$ programs, a fuzzer, and $c$ available CPU cores, the objective is to design a scheduler that continuously and dynamically allocates and deallocates the fuzzing of the programs among the cores, so as to optimize the code coverage across all programs. This task requires making informed scheduling decisions while accounting for the unpredictable nature of fuzzing, the diversity of software, and the inherent constraints of limited computational resources, i.e. limited amount of available CPUs. 
It is important to emphasize that the programs are fuzzed independently, utilizing the same off-the-shelf fuzzer that  lacks any specialized functionality for conducting fuzz testing on multiple programs simultaneously. We use the term \emph{fuzzers}, to refer to the instances of the same fuzzer running on different programs. Furthermore, sometimes we use the term \emph{program} or \emph{fuzzed program} to refer to the concept that the fuzzer is fuzz testing the program.

Central to fuzzing at scale is the scheduler. However, prior to delving into it, it is important that we quickly touch on two other aspects.

\vspace{10pt}
\noindent\textbf{Condition $p > c$.} The fuzzing at scale problem is truly challenging, when the number of programs $p$  exceeds the number of available CPU cores $c$. When this condition is met, the scheduler must make intelligent decisions regarding which programs to run and when to allocate the limited CPU resources. If the number of programs is comparable or less than the number of cores, all programs can be fuzzed  simultaneously without the need for a scheduler to make any decisions, thus eliminating the challenge of resource allocation and making the problem trivial. 

\vspace{10pt}
\noindent\textbf{Code coverage objective.}
It is essential to assume that the fuzzer's objective aligns with the fuzzing at scale objective, i.e., both should be focused on maximizing code coverage (or any other shared objective, such as the number of found bugs). 
If there is a misalignment in the objectives, then the scheduler will be either inefficient or entirely meaningless. Fortunately, for many existing fuzzers, this alignment is already in place, as their primary aim is to uncover as many unique program code regions as possible, i.e. to maximize code coverage.
The objective of the fuzzing at scale thus becomes to schedule fuzzing of some programs $P_1,\ldots,P_p$ so as to maximize the accumulative coverage over all these programs, i.e. 
\begin{equation}
\label{eq:max}
max \sum\limits_i Cov (P_i),
\end{equation}
where $Cov(P_i)$ is the total code coverage obtained by fuzzing the program $P_i$, while $\max$ is taken over all schedulers.

\subsection{Schedulers}
\label{sect:scheduler}

A scheduler plays a pivotal role in fuzzing at scale tasks, serving as the primary mechanism for  simultaneously fuzzing programs across available CPU cores. Similarly to the objectives of operating system schedulers~\cite{peterson1985operating}, the goal of a fuzzing scheduler is to allocate fuzzing programs on the available CPU cores, continuously rotating which programs are actively fuzzed based on a predefined criterion. By managing the execution of fuzzing tasks, the ultimate goal of the scheduler is to ensure that the available CPU cores are utilized optimally to maximize code coverage across the entire pool of programs while adhering to the limited CPU cores constraint.

We tackle the scheduling problem with multitasking. 
In other words, all programs are fuzzed simultaneously, but due to the limited number of CPUs, the fuzzers take turns to run on the available CPUs.
More precisely, we divide the available CPU time into short, equal periods called \emph{time slices}. At each time slice only $c$ among all $p$ programs are actually being fuzzed on the available $c$ cores. The remaining fuzzers are paused -- in  operating system terminology they are in so-called stopped state, so we will use the term \emph{stopped} to refer to fuzzers that have been paused. 
At the beginning of a time slice, the scheduler decides which stopped fuzzer to start (resume) running. 
To further simplify the decision, we always stop the longest continuously running fuzzer. It means that if the fuzzing of a program starts at time slice i, it will be stopped at i + c. 
As a result, \emph{the only task of a scheduler is at each time slice to choose which stopped fuzzer  to start}. 
During the entire fuzzing at scale process, each program fuzzer is started and stopped multiple times.  

Let us now consider potential schedulers, beginning with the basic (or baseline) and moving to more advanced.  
As a baseline we use a simple scheduler built upon the round-robin scheduling algorithm~\cite{liu1973scheduling} (in operating systems this scheduler is also considered basic).
At each time slice, it chooses to start the fuzzer that has not been ran for the longest of time. This is equivalent to cycling through the pool of programs. 
By doing so, the round-robin scheduler aims to distribute the fuzzing efforts evenly across all programs. 
Clearly, this baseline scheduler is not  optimized for maximizing code coverage, i.e. it is not necessarily efficient, because it does not have any mechanism to give preference to certain fuzzers.

The next logical step in developing a more efficient scheduler is to incorporate feedback from the fuzzed programs, to make informed scheduling decisions, rather than to rely solely on their order in the pool. 
This feedback-driven approach allows the scheduler to prioritize programs with promising fuzzing results. It leads to a more efficient and effective allocation of CPU cores.
As a feedback, we use the code coverage obtained by the fuzzer per unit of fuzzed time.  Furthermore, we assume that this past code coverage is a predictor for future code coverage. In other words, if in the past, fuzzing program A resulted in more code coverage then fuzzing program B, then most likely this will be the case in the future as well. 
All of our schedulers will rely on this hypothesis and thus will discriminate in favor of fuzzing programs that have resulted in higher past coverage.

To incorporate the code coverage feedback from the fuzzed programs, we employ the multi-armed bandits (MAB) algorithm~\cite{slivkins2019introduction}, a well-established decision-making approach designed to optimize long-term rewards, which in our fuzzing at scale context corresponds to maximizing code coverage (\ref{eq:max}). We assume that at the beginning of each time slice, the scheduler 
must decide which 
stopped (paused) program should resume fuzzing. The MAB algorithm guides this decision-making process by considering the previous performance of each program, i.e. by factoring 
the amount of code coverage achieved per unit of time during a program fuzzing.
\begin{algorithm}[tb]%
\caption{Simple MAB Scheduler\protect\\
\small Pseudo-code for the Simple MAB Scheduler. Each fuzzed program $P_i$ as assigned bandit, defined with achieved coverage $Cov_i$, fuzz time $Time_i$, and reward $\frac{Cov_i}{Time_i}$. Lists $L_{run}, L_{stop}$ hold indices of programs that are currently undergoing fuzzing and paused, respectively. $getCov(P_i, t-1)$ is a function that returns the increase of code coverage of fuzzing program $P_i$ during the time slice $t-1$. $RandFloat(0,1)$ returns random number in the interval $(0,1)$. $RandElemFrom(L_{stop})$ returns random element from the list $L_{stop}$.}
\label{alg:simplemab}
\DontPrintSemicolon
\SetAlgoLined
\KwIn{$P_1,\ldots,P_p$ programs, fuzzer $F$, $c$ CPU cores, $\epsilon$}
\For{$i=1,\ldots p$}{
    $Cov_i= Time_i=0$ \comm{init MABs}
}
$L_{run} = \{1,\ldots,c\}$                      \hspace{70pt} \comm{init list running}\;
$L_{stop} = \{c+1,\ldots,p\}$
\hspace{45pt} \comm{init list stopped}\;
\For{ $i$ in $L_{run}$} {
    START $F(P_i)$ \comm{start fuzzing programs from $L_{run}$}
}
\comz{At beginning of each time slice }\\
\For{ $t=1,\ldots$ }{
    \comz{Update MABs} \\
    \For{ $i$ in $L_{run}$}{
        $Time_i\ += 1$ \comm{add time}\\
        $Cov_i += getCov(P_i, t-1)$  \comm{add new coverage produced in the prev time slice}
    }
    $last = L_{run}[c]$ \comm{get last running} \\
    STOP $F(P_{last})$ \comm{stop fuzzing it} \\
    $L_{stop} = L_{stop} \cup \{last\}$ \\
    $L_{run} = L_{run} \setminus \{last\}$ \\
    \comz{Use MAB to decide which to fuzz next} \\
    \eIf{ $\epsilon < RandFloat(0,1)$}{
        $new = \arg\max\limits_{i \in L_{stop}} \frac{Cov_i}{Time_i}$  \comm{get best program}
    }{
        $new = RandElemFrom(L_{stop})$ \comm{get random}
    }
    START $F(P_{new})$ \comm{start/resume fuzzing it} \\
    $L_{stop} = L_{stop} \setminus \{new\}$ \\
    $L_{run} = \{new\} \cup L_{run}$ \\
}
\end{algorithm}
\noindent
It does not always choose the best-performing fuzzer, as that is not an optimal strategy in the long run. 
Rather, in accordance to the MAB approach, the scheduler dynamically balances exploration (choose random fuzzer) and exploitation (choose best performing fuzzer), ensuring that it maintains a diverse pool of fuzzing targets while also focusing on programs with the highest potential for  code coverage. The balance (or the tradeoff) between exploration and exploitation is controlled by a parameter $\epsilon$ where $0<\epsilon<1$ and is called $\epsilon$-greedy~\cite{sutton2018reinforcement}. For example, when $\epsilon=0.1$, then in 10\% of the time slices MAB will schedule for fuzzing a randomly chosen program, and in 90\% the best performing program (i.e. the one that achieved highest coverage per time unit so far). We call this scheduler Simple MAB scheduler and give its pseudo code in Algorithm~\ref{alg:simplemab}. 

\begin{figure}[h]
    \centering
    \includegraphics[scale=0.2]{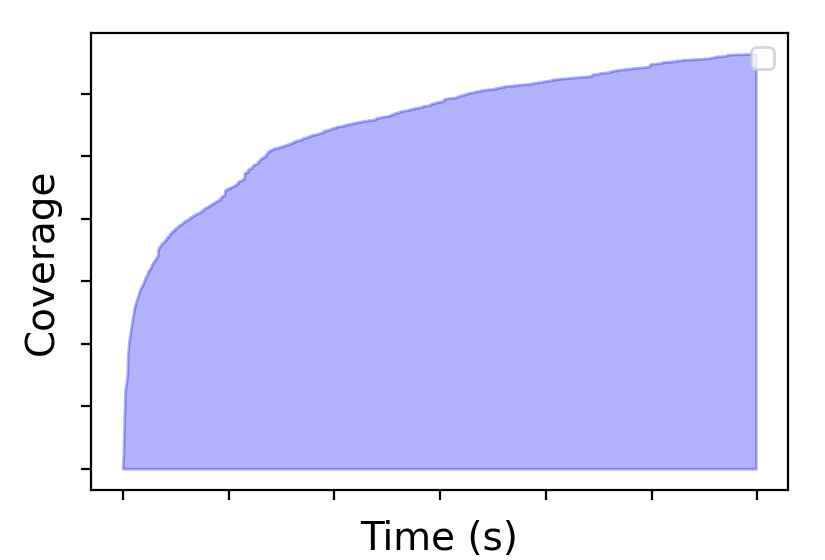}
    \caption{The total amount of code coverage over time obtained by fuzzing  1,000 programs with  AFL++ fuzzer.}
    \label{fig:code-coverage}
\end{figure}

Multi-armed bandits are effective under the so-called \emph{stationary} assumption, which posits that the rewards remains constant. In our case, this translates to the assumption that each program gets a constant increase in coverage during fuzzing. However, in practice this is not true. Observations indicate that for the majority of programs, fuzzers provide more code coverage at the beginning of fuzzing, followed by a decline -- our own experiments confirm this finding as well, refer to Figure~\ref{fig:code-coverage}. 

This outcome is not unexpected, as not all so-called program branches (i.e. conditional statements \code{if, switch}) can be easily flipped (e.g. force statement in \code{if} to change value from true to false and vice-versa). Instead, the simpler branches are flipped early on with minimal effort, while substantial time is devoted to tackling the more complex ones. 
We thus next investigate MABs  featuring \emph{non-stationary} rewards, which are characterized by rewards that vary over time, or in our context, a code coverage rate that declines with time.
To handle this case, we apply discounting~\cite{sutton2018reinforcement}. Specifically, instead of presuming that all code coverage contributes equally to reward calculations, we exponentially discount earlier coverage results using a factor $\gamma$, where $0<\gamma<1$. More precisely, we update the reward estimate for a program from  the old $\frac{\sum_{i=1}^{n} c_i}{\sum_{i=1}^{n} t_i}$, where $c_i$ and $t_i$ represent the corresponding code coverages and time intervals it ran, to the new reward $\frac{\sum_{i=1}^{n} c_i \cdot \gamma^{n-i} }{\sum_{i=1}^{n} t_i \gamma^{n-i}}$. This  signifies that we assign exponentially greater importance to recent fuzzing results compared to older outcomes. We refer to this scheduler as the Discounted MAB scheduler.

\begin{algorithm}[tb]%
\caption{\tool Scheduler\protect\\
\small Pseudo-code for \tool scheduler. $timeToIncreaseEpsilon(t)$, $timeToChangeGamma(t)$ are predicates. For other notations refer to Algorithm~\ref{alg:simplemab}.}
\label{alg:final}
\DontPrintSemicolon
\SetAlgoLined
\KwIn{$P_1,\ldots,P_p$ programs, fuzzer $F$, $c$ CPU cores, $\epsilon$}
\For{$i=1,\ldots p$}{
    $LTime_i = LCov_i=\emptyset$ \comm{init MABs lists}
}
$L_{run} = \{1,\ldots,c\}$                      \hspace{70pt} \comm{init list running}\;
$L_{stop} = \{c+1,\ldots,p\}$
\hspace{45pt} \comm{init list stopped}\;
\For{ $i$ in $L_{run}$} {
    START $F(P_i)$ \comm{start fuzzing programs from $L_{run}$}
}
$\epsilon = \epsilon_{min}$ \comm{starting $\epsilon$ value} \\
$\Gamma = \{\gamma_1,\ldots, \gamma_g\}$ \comm{all $\gamma$ values}  \\
$\gamma = \gamma_1$ \comm{starting $\gamma$ value} \\
\comz{At beginning of each time slice }\\
\For{ $t=1,\ldots$ }{
    \comz{Update MABs} \\
    \For{ $i$ in $L_{run}$}{
        Append $1$ to $LTime_i$ \comm{append new time} \\
        Append $getCov(P_i, t-1)$ to $LCov_i$  \comm{append new coverage}    \\
    }
    $last = L_{run}[c]$ \comm{get last running} \\
    STOP $F(P_{last})$ \comm{stop fuzzing it} \\
    $L_{stop} = L_{stop} \cup \{last\}$ \\
    $L_{run} = L_{run} \setminus \{last\}$ \\
    \comz{Update $\epsilon,\gamma$ if right time step } \\
    \If{ $timeToIncreaseEpsilon(t)$ }{
        $\epsilon = increase(\epsilon)$
    }
    \If{ $timeToChangeGamma(t)$ }{
        $\gamma = next(\Gamma, \gamma)$
    }
    \comz{Use MAB to decide which to fuzz next} \\
    \eIf{ $\epsilon < RandFloat(0,1)$}{
        $new = \arg\max\limits_{i \in L_{stop}}  \frac{\sum_j LCov_i[j] \cdot \gamma^{n-j} }{\sum_j LTime_i[j] \cdot \gamma^{n-j}}$  
        \comm{get best prog}
    }{
        $new = RandElemFrom(L_{stop})$ \comm{get random}
    }
    START $F(P_{new})$ \comm{start/resume fuzzing it} \\
    $L_{stop} = L_{stop} \setminus \{new\}$ \\
    $L_{run} = \{new\} \cup L_{run}$ \\
}
\end{algorithm}

In the present MAB framework, we must determine the values for two parameters: the exploration-exploitation balance $\epsilon$ and the discount rate $\gamma$. Identifying the optimal values for these parameters can be challenging or even unattainable. As a result, we employ a more cautious approach, allowing these values to fluctuate rather than remain fixed. Specifically, we permit $\gamma$ to continually cycle through a range of values. This implies that, at times, we assume older code coverage results are more significant, while at other times, they are deemed less. Conversely, we gradually increase the value of $\epsilon$ over time. This strategy initially compels the scheduler to allocate fuzzing resources to programs that demonstrate the largest increase in code coverage. 
However, as time progresses, the scheduler explores more other programs as well. 
The rationale behind this approach is that, over time, previous code coverage becomes a less reliable predictor of future coverage. This is because only challenging, unflipped program branches are likely to remain, and they are equally hard to flip for all programs, regardless of the previous code coverage progress.
Thus, by gradually increasing $\epsilon$, we assure the scheduler is giving more balanced fuzz time to all programs. However, we never increase the value of $\epsilon$ to 1, so the scheduler still favors, albeit less, programs that have achieved higher code coverage. 
We refer to our final scheduler as \tool, and present its pseudo-code in  Algorithm~\ref{alg:final}.

In Table~\ref{tbl:schedulers:brief}, we provide a brief overview of the examined schedulers, outlining the problems they aim to address and the solutions they propose.

\vspace{10pt}
\noindent
\textbf{Measure of efficiency. }
Assessing the efficiency of a scheduler can be ambiguous. In this context, we explore two distinct metrics of efficiency.
The first metric focuses solely on the accumulative coverage across all fuzzed programs. This equates to the total sum of all coverages of individual programs.
It is aligned with the objective of the scheduler. 
This metric referred to as \emph{accumulative} is relevant when the aim is to measure total coverage. 
However, it neglects the possibility that a smaller subset of programs may contribute a disproportionately large portion to the overall coverage. In essence, it measures the mean code coverage across all programs but ignores the variance. 

The second metric, referred to as \emph{voting}, attempts to correct this potential oversight by considering the variance. It compares fuzzing outcomes based on the fraction of programs that have been fuzzed better. For instance, in a scenario where 100 programs are fuzzed, if approach A yields better coverage than approach B for 35 programs, worse coverage for 25, and same coverage for the remaining 40, then according to the second  metric (voting), approach A would be deemed superior because $35>25$. 
(Furthermore, we say that A is 10\% better than B because $\frac{35-25}{100} = 0.1$.) 
Conversely, the first metric (accumulative) would compute the total coverage produced for all 100 programs and decide on the preferable approach solely based on this sum.
Therefore, the accumulative metric prioritizes candidates that yield the highest overall aggregate coverage, whereas the vote metric gives preference to those providing higher coverage for the majority of the programs.

\begin{table*}[htbp]
\caption{Schedulers for fuzzing at scale. 
Each subsequent \textbf{Scheduler} aims to address a remaining \textbf{Problem} that the previous scheduler left unsolved, by utilizing the \textbf{Solution}.
}
\label{tbl:schedulers:brief}
\renewcommand{\arraystretch}{1.3}
\centering
\begin{tabular}{|l|p{4cm}|p{5cm}|}
\hline
\textbf{Scheduler} & \textbf{Problem} & \textbf{Solution}\\
\hline
\hline
Baseline (round-robin) & basic scheduling of multiple fuzzers & equal time to all fuzzers; in each time slice choose the longest paused fuzzer
\\
\hline
Simple MAB & 
prefer high code coverage fuzzers & 
MAB; compute rewards as coverage/time and either choose fuzzer with best reward (in $1-\epsilon$ of time slices) or random fuzzer (in $\epsilon$ of time slices)
\\
\hline
Discounted MAB & 
prefer more recent high code coverage fuzzers & 
discounted MAB; when computing rewards give exponentially lower importance to older coverage results (by multiplying each $T$-periods old with  $\gamma^T$) 
\\
\hline
\tool & 
choose good discounting $\gamma$; with time, past coverage becomes less reliable predictor of future coverage &
discounted MAB but cycle $\gamma$ and slowly keep increasing $\epsilon$
\\
\hline
\end{tabular}
\end{table*}

\subsection{Fuzzing at Scale}

We can use each of the schedulers described in the previous section to simultaneously fuzz test multiple programs and thus carry out fuzzing at scale. 
Recall, a common feature of all these schedulers is the idea of multitasking, i.e. frequently switching between executing fuzzers on the available processing cores. This switching process occurs once at the beginning of each time slice and the choice on which fuzzer to resume executing is determined by the scheduler. Further we provide additional details about the internals of the framework:
\begin{itemize}
\item 
\textbf{Time slice duration.} 
Shorter time slices are preferable because they result in finer granularity, 
leading to reduced time spent on underperforming fuzzers.
However, if the slices are too short, 
the impact of recomputing the MAB rewards and starting or stopping fuzzers may demand significant CPU resources.
Furthermore,  the duration of the time slices should be sufficiently large to facilitate proper support and implementation by the operating system (OS), which also relies on a scheduler for multitasking purposes.
This implies that the fuzzing scheduler's time slices must exceed those employed by the OS scheduler, which, in the case of the present Linux kernel, default to 20 ms. 
Otherwise, the OS scheduler cannot perform the context switch on time, i.e. cannot preempt one program and resume another. 
\item 
\textbf{Fuzzing at scale on clusters. } If the time slices are sufficiently long, 
the aforementioned multitasking approach can be used to fuzz at scale, not only on an individual machine, but also across a cluster. 
The scheduler will run on the server (in a similar fashion as in the case of a single machine), but will communicate with fuzzers that run on client machines. To accommodate this scenario, the slice duration should be set larger than the latency between the nodes of the cluster.  %
If this condition is met, our framework can be easily extended to support multiple machines.
\item
\textbf{Limit on the number of programs.} The OS may impose limit on the number of  programs that can be fuzzed concurrently.  
This restriction aligns with the standard limitations of multitasking OSs, which involve managing multiple processes that consume varying amounts of system resources, including RAM, disk usage, open files, etc.
Based on our experimental results presented in Section~\ref{sect:eval}, the imposed limit appears to be relatively generous, as we were able to fuzz up to 5,000 programs simultaneously on a single Linux machine.

\end{itemize}

\section{Benchmarks}
\label{sect:ubuntubench}

We now shift our focus to the second major component of a fuzzing at scale framework: a substantial collection of target programs for fuzzing. 
Most of the currently used benchmarks for fuzzers are rather small. We inspected all such collections presented in ~\cite{koike2022slopt,nikolic2021refined,gan2018collafl,aschermann2019redqueen,osterlund2020parmesan,yun2018qsym,rawat2017vuzzer,chen2018angora,liang2022pata,lyu2019mopt,lyu2022ems,liang2019deepfuzzer,lemieux2018fairfuzz,chen2019enfuzz,yue2020ecofuzz,she2022effective,geretto2022snappy,yu2022htfuzz,lee2021constraint,chen2020muzz,gan2020greyone,zong2020fuzzguard,nguyen2020binary,chen2020meuzz,nagy2021breaking,zhou2022ferry,wangcarpetfuzz,blazytko2019grimoire,shah2022mc2,zhu2021regression,nagy2021same,chen2019matryoshka,cho2019intriguer} and found that they consist of 10-30 programs, and in total, across all of these 33 papers, there are only 184 distinct programs. As a general rule, these programs are sampled from the \ossfuzz~\cite{serebryany2017oss} package -- a curated collection of open-source programs and libraries. Currently, \ossfuzz contains around 1,000 target programs, but only roughly half of them have source code in C/C++. 
Furthermore, due to widespread adoption as a predominant fuzzing target, there is a risk of overfitting. That is, recent fuzzers might inadvertently be optimized to excel on programs from \ossfuzz, while only performing adequately on other targets. Therefore, using \ossfuzz as a sole  benchmark may not provide objective evaluation results.

This calls for the development of a new fuzzing benchmark that will include a large number of programs. Such a set is significant for both the success of our fuzzing at scale task and for the advancement of future fuzzing research. 
A comprehensive and diverse benchmark offers numerous benefits for fuzzing projects. It provides a rich and representative sample of programs, capturing the complexity and variety of real-world software. This enables the evaluation of fuzzing techniques in a more realistic context.
The task of creating benchmark includes identifying potential candidate programs, retrieval of their corresponding compilation rules, determining the appropriate invocation methods for input execution, and providing testcases.

The currently limited number of candidate programs might be attributed to the fact that benchmarks are created manually. That is a tedious and time-consuming process. 
In addition, maintaining and updating a manually produced benchmark becomes an ongoing task, as otherwise, the reuse of the same benchmarks for prolonged period does not take into account that software is continuously changing.  
Producing the benchmark automatically is a valuable alternative because it is the most efficient and scalable way to generate a large and diverse set of target programs. 
Moreover, automating the process ensures easy rotation of programs in the benchmark, as well as staying up-to-date with rapidly evolving software. 
In comparison to manual production, there are a few drawbacks such as the potential exclusion of certain important programs or the inclusion of irrelevant. However, we believe the tradeoff is well justified because the advantage of having a large, diverse, and up-to-date benchmark outweights the potential drawbacks of not including certain programs. Thus we focus on generating the benchmark automatically. 

To obtain programs for inclusion in the benchmark, we opt to utilize the set of Ubuntu packages as our primary source. 
This offers multiple benefits. First, it provides a single, unified source consisting of a vast pool of programs, simplifying the process of collecting, maintaining, and even updating the benchmark. Second, the programs within Ubuntu packages are highly diverse, spanning various application domains, thus ensuring that the benchmark is representative and well-suited for evaluating fuzzing techniques across a wide range of software. Third, a considerable number of these programs adhere to an installation template,
which simplifies the process of setting up and incorporating programs into the benchmark. 

We begin by considering the entire set of available Ubuntu packages and apply several \emph{fully automatic} steps to obtain the final benchmark:
\begin{enumerate}

\item 
\textbf{Eliminate unsuitable candidates.} We remove obviously unsuitable candidates based solely on the names of the packages. For instance, packages with names beginning with \textsf{lib\*} are excluded, as libraries typically require manual inclusion in larger programs, rendering them unsuitable for automatic benchmark generation.

\item  
\textbf{Compile remaining programs.} We attempt to compile all remaining programs, either by utilizing the provided \textsf{makefile}s or by resorting to general compilation routines when necessary.

\item 
\textbf{Identify executable binaries and inputs arguments.} We examine all compiled packages and identify executable binaries. For each such program, we apply known heuristics (as indicated in Section~\ref{sect:implementation}) and try to infer 
input arguments that would prompt reading of files, so the fuzzer can supply testcases during fuzz testing the program. 

\end{enumerate}

All such candidate programs are collected to form the fuzzing benchmark \benchmark.

\section{Implementation}
\label{sect:implementation}

We implement the scheduler \tool in C++ for efficiency purposes, comprising 4,320 lines of code. 
The same implementation can be utilized to run the other aforementioned schedulers, by disabling some of the functionalities of \tool, as all of these schedulers are subsets of \tool. 
Internally, the scheduler maintains two lists of fuzzed programs: those that are currently running and those that are stopped. Each fuzzed program is executed as a separate process, running the fuzzer on the target program.
At the beginning of each time slice, the scheduler makes a decision on which stopped fuzzed program to start, as previously mentioned, based on the multi-armed bandits  algorithm. The actual stopping and starting of fuzzers are accomplished by sending the signals \code{SIGSTOP} and \code{SIGCONT} to the corresponding processes running the fuzzer on programs.
\tool enables the selection of the number of utilized CPU cores by setting hard affinity~\cite{peterson1985operating} for all processes initiated by the scheduler. This is achieved through a system call implemented with the C function \code{sched\_setaffinity}, which allows to select precisely the set of CPU cores allowed to run the fuzzing processes. 
\tool also incorporates a range of mechanisms that mitigate the excessive utilization of CPU and RAM resources by the targeted programs by continuously monitoring and potentially terminating processes with the use of the library \code{libprocps}. 
This feature is particularly useful, given that the \benchmark is automatically compiled and may potentially include resource-intensive programs.

We implement the generation of \benchmark as a series of Python scripts, one for each step of this process, and in total with 1965 lines of code. All of these scripts support parallelization by utilizing the Python module \textsf{multiprocessing}. We first download all Ubuntu package that pass the filtering criterion based on their name. Then, before compiling the packages we install the necessary dependencies according to the information provided in the file  \code{debian/control} that is present in each Ubuntu package. 
We try to compile the Ubuntu packages with dedicated C/C++ compilers provided by the fuzzers, so that the produced binaries are well instrumented to provide code coverage information. 
For this purpose, we initially employ the prevalent method of executing \code{debian/rules build} that is also present in all packages. 
When this approach does not generate an instrumented binary, we resort to an ad-hoc compilation process, based on installation files such as \code{autogen.sh, Makefile,  CMakeLists.txt}, which indicate the appropriate build system to use.
By leveraging these files, we can specify to use fuzzer provided compilers and obtain instrumented binaries. 
To detect the required fuzzer instrumentation in compiled binaries, we search for an appropriate fuzzer fingerprint. This is done by parsing the output produced with the command line tool \code{objdump}.
We also check for dependencies of the binaries on dynamic libraries and on particular requirement for currently working directory, which is important to binaries that encode relative path. 
In order to deduce the input arguments that prompt the binary to read from an input file, we employ a heuristic approach and parse the outputs of \code{./binary –help} and \code{./binary - -help}, and subsequently examine all plausible candidates. To determine whether the binary indeed reads from a file when a certain input argument is provided, we utilize the command line tool \code{strace}, which 
detects file-opening operations for reading. 
This method enables us to effectively identify specific input arguments that cause the binary to perform file-reading actions.

All of the above implementations are publicly released~\footnote{\url{https://github.com/ivicanikolicsg/fuzzing\_at\_scale}.}.

\section{Evaluation}
\label{sect:eval}

Further we present experimental results highlighting the 
importance of schedulers and the relevance of the whole fuzzing at scale framework. 

\vspace{10pt}
\noindent\textbf{Experimental setup.}
For all experiments we use the same box with 
Ubuntu Desktop 20.04, two Intel Xeon E5-2680v2 CPUs @2.80GHz with total of 40 cores, and 64GB DDR3 RAM. 
All subsequent fuzzing at scale tasks are executed on 30 cores and all the schedulers are set to use 0.1s times slices. 
In these tasks, the number of fuzzed programs and the duration of fuzzing are determined by two basic goals: provide meaningful results and  finish all experiments within two weeks. 
For these reasons, the number of fuzzed programs ranges in 1,000 to 5,000 for the larger benchmark (\benchmark) and around 200 to 300 for the smaller (\ossfuzz), and the fuzzing duration in 15 minutes to 1 hour on average per program. 
For the actual fuzzing, we make use of  three well-established fuzzers, AFL~\cite{zalewski2017american},  AFL++~\cite{fioraldi2020afl} and Honggfuzz~\cite{swiecki2021honggfuzz}. 
This selection of fuzzers is based on the fact that AFL is the most popular baseline grey-box fuzzer, whereas AFL++ and Honggfuzz are the top performing fuzzers on Google fuzzer comparison platform FuzzBench~\cite{metzman2021fuzzbench}.
As a measure of code coverage we use edge coverage provided by AFL  tool \code{afl-showmap}. 

Instead of running each experiment multiple times and reporting the average of those runs, we run each only once and report the outcome. We do this for two reasons. First, the number of experiments in Sections~\ref{sect:eval:compare-schedules}, \ref{sect:eval:scheduler-vs-fuzzer} alone is 14, each runs for around half a day, thus we need close to a week just to run once all experiments. Multiple runs would thus require a month or so of running. Second, the large number of programs in each experiment significantly reduces the variance between different runs. Therefore, repeating experiments and averaging their outcomes may not significantly improve the accuracy of the reported results.

\vspace{10pt}
\noindent\textbf{The benchmarks \benchmark and \ossfuzz.}
As described in Section~\ref{sect:ubuntubench}, we use packages from Ubuntu to obtain the benchmark \benchmark -- refer to Table~\ref{tbl:benchmark} for the steps. More precisely, from the pool of all 75,799 available packages from Ubuntu 20.04, %
with filtering and compilation we produce 5,467 target programs that compose \benchmark. All of these programs can be compiled at least with the standard \code{gcc} and \code{g++} compilers. 
During fuzzing, as initial testcases (seeds) we provide randomly chosen files found in the corresponding Ubuntu packages.  
In the \ossfuzz suite~\cite{serebryany2017oss}, out of the total 1,193 packages, a mere 326 possess C/C++ source code and are recognized as being compatible with the AFL++ compiler, as indicated in their respective configuration files. These yield a total of 345 distinct binaries that constitute our \ossfuzz benchmark set. Out of these binaries, 216 can also be compiled with AFL compiler.

\begin{table}[h]
\caption{The number of  packages/programs handled at different stages of producing \benchmark. }
\label{tbl:benchmark}
\centering
\renewcommand{\arraystretch}{1.3}
\setlength\tabcolsep{10pt}
\begin{tabular}{|l||r|r|}
\hline
 & \benchmark & \ossfuzz
 \\
\hline
\hline

Targets  packages           &  75,799 & 1,193\\
Filtered packages           &  22,368 & 326 \\
Compiled packages           &  15,551 & 326 \\
\textbf{Final binaries}     &  \textbf{\ 5,467} & \textbf{345}  \\
\hline
\end{tabular}
\end{table}

\subsection{The Importance of a Scheduler}
\label{sect:eval:compare-schedules}

Our first experiment is to evaluate the significance of schedulers in fuzzing at scale tasks. More precisely, we want to test empirically if the choice of scheduler can impact the outcome (produced coverage) of this task. For this purpose we compare all schedulers discussed in Section~\ref{sect:scheduler}. 
We consider 4 different schedulers: baseline round-robin, simple MAB, discounted MAB with fixed $\gamma=0.9$\footnote{The value of $0.9$ was taken as the best performing among our tested set $\{0.9, 0.99, 0.999\}$.}, and \tool with cycling $\gamma \in \{0.9, 0.99, 0.999\}$ and $\epsilon$ increasing in $[0.01,0.75]$. 
For each scheduler, we conduct experiments independently but use the same set of selected programs. The programs are fuzzed simultaneously with AFL++ on 30 CPUs. 
We run two fuzzing at scale tasks: one for \benchmark with 1,000 programs sampled at random and with average fuzz time of 15 minutes, and one for \ossfuzz where we take all available 345 programs and fuzz for 1 hour on average per program\footnote{It means  that each experiment for both \benchmark and \ossfuzz runs roughly for half a day.}.
We evaluate the performance of the four schedulers on both metrics, accumulative and voting. Recall, accumulative metric favors schedulers that provide better total coverage across all fuzzed programs, whereas voting metric favors those that provide higher coverage for more programs.
We present the results as pairwise comparisons of the different schedulers and in terms of percentage increase or decrease between the first and the second elements of the pair. 
\definecolor{colorplus}{rgb}{0.741, 0.906, 0.741}
\definecolor{colorminus}{rgb}{1.0, 0.714, 0.702}

\begin{table}[h]
\caption{Comparison of 4 different schedulers with accumulative metric. The numbers in the cells are the percentage increase/decrease in total coverage provided by the row scheduler in comparison to the column scheduler. 
}
\label{tbl:sched:accum}
\renewcommand{\arraystretch}{1.3}
\setlength\tabcolsep{6pt}
\begin{tabular}{|l|cccc|cccc|}
\multicolumn{1}{c}{} & \multicolumn{4}{c}{\benchmark} & \multicolumn{4}{c}{\ossfuzz} \\
\hline
& 
\rotatebox{90}{Baseline } &  \rotatebox{90}{Simple } & 
\rotatebox{90}{Disc } &  \rotatebox{90}{\tool } 
& 
\rotatebox{90}{Baseline } &  \rotatebox{90}{Simple } & 
\rotatebox{90}{Discounted } &  \rotatebox{90}{\tool } \\
\hline
Baseline & 
0 & \cellcolor{colorminus}-13 & \cellcolor{colorminus}-16 & \cellcolor{colorminus}-22 & 
0 & \cellcolor{colorminus}-10 & \cellcolor{colorminus}-8 & \cellcolor{colorminus}-9
\\
Simple MAB & 
\cellcolor{colorplus}15 & 0 & \cellcolor{colorminus}-3 &  \cellcolor{colorminus}-10 & 
\cellcolor{colorplus}11 & 0 & \cellcolor{colorplus}1 &  0 
\\
Discounted & 
\cellcolor{colorplus}19 & \cellcolor{colorplus}3 & 0 &  \cellcolor{colorminus}-7 &
\cellcolor{colorplus}9 & \cellcolor{colorminus}-1 & 0 & 0
\\
\tool & 
\cellcolor{colorplus}29 & \cellcolor{colorplus}11 & \cellcolor{colorplus}8 & 0 &
\cellcolor{colorplus}10 & 0 & 0 & 0
\\
\hline
\end{tabular}%

\end{table}

Let us first take a look at the results in accumulative metric given in Table~\ref{tbl:sched:accum}.
First, it is clear that the baseline is the worst performing scheduler.
This indicates that, when fuzzing simultaneously multiple programs (fuzzing at scale), allocating equal fuzz time to each program (a strategy implemented by the baseline scheduler via simple round-robin allocation) provides inferior accumulative code coverage over all programs. All other considered schedulers provide up to 10\%-29\% higher code coverage in comparison to the baseline. In particular, 
according to the results of Table~\ref{tbl:sched:accum}, the top scheduler on \benchmark is \tool which provides clear advantage over the remaining schedulers. 
Similar results are obtained on \ossfuzz benchmark as well --  \tool is the top performing scheduler.
From the fuzzing logs provided by our implementation we could also get information about the exact fuzzing time allocated to each program. 
Accordingly, the baseline scheduler assigned approximately 15 minutes (900 seconds) to fuzzing of each program.  In contrast, the time assigned by other schedulers varies. For instance, \tool assigned from 119 seconds to
8,878 seconds for fuzzing of different \benchmark programs, and 390 to 8,026 seconds for \ossfuzz programs. Thus, by monitoring and adjusting the allocated fuzzing time, it provided 10-29\% higher total coverage in comparison to the baseline.

\definecolor{colorplus}{HTML}{BDe7BD} 
\definecolor{colorminus}{HTML}{FFB6B3}

\begin{table}
\caption{Comparison of 5 different schedulers with the voting metric. The numbers in the cells are the percentage increase/decrease of fuzzed programs for which the row scheduler provides better coverage than the column scheduler. 
}
\label{tbl:sched:vote}
\renewcommand{\arraystretch}{1.3}
\setlength\tabcolsep{6pt}
\begin{tabular}{|l|cccc|cccc|}
\multicolumn{1}{c}{} & \multicolumn{4}{c}{\benchmark} & \multicolumn{4}{c}{\ossfuzz} \\
\hline
& 
\rotatebox{90}{Baseline } &  \rotatebox{90}{Simple } & 
\rotatebox{90}{Disc } &  \rotatebox{90}{\tool } 
& 
\rotatebox{90}{Baseline } &  \rotatebox{90}{Simple } & 
\rotatebox{90}{Discounted } &  \rotatebox{90}{\tool } \\
\hline

Baseline & 
0 & \cellcolor{colorplus}47 & \cellcolor{colorplus}19 &  \cellcolor{colorminus}-8 &
0 & \cellcolor{colorplus}17 & \cellcolor{colorplus}3 &  \cellcolor{colorminus}-9
\\
Simple MAB & 
\cellcolor{colorminus}-47 & 0 & \cellcolor{colorminus}-45 &  \cellcolor{colorminus}-56 &
\cellcolor{colorminus}-17 & 0 & \cellcolor{colorminus}-11 &  \cellcolor{colorminus}-21
\\
Discounted & 
\cellcolor{colorminus}-19 & \cellcolor{colorplus}45 & 0 &  \cellcolor{colorminus}-24 &
\cellcolor{colorminus}-3 & \cellcolor{colorplus}11 & 0 & \cellcolor{colorminus}-9 
\\
\tool & 
\cellcolor{colorplus}8 & \cellcolor{colorplus}56 & \cellcolor{colorplus}24 &  0 &
\cellcolor{colorplus}9 & \cellcolor{colorplus}21 & \cellcolor{colorplus}9 &  0
\\
\hline
\end{tabular}%
\end{table}

We now turn our attention to the results of the voting metric, presented in Table~\ref{tbl:sched:vote}. Several key differences from the previous metric are worth noting, the most striking one being that the baseline scheduler demonstrates good performance in this metric as it surpasses the performance of some of the other schedulers.
It means allocating equal fuzz time to each program is a reasonable approach when aiming to generate good coverage for most of fuzzed programs. However, even under this metric, there is a scheduler that strictly outperforms the baseline, and this is \tool.
This new scheduler provides better coverage in comparison to the baseline, as well as the other two schedulers. 

From the analysis presented above, we can conclude  that the choice of a scheduler significantly influences the outcome of concurrently fuzzing multiple programs from the two distinct benchmarks, 
as demonstrated by the two different code coverage metrics. 
Specifically, employing a more sophisticated scheduler like \tool can yield up to around 30\% greater fuzzing efficiency (based on the metrics) as opposed to simply allocating equal fuzzing time to each program.
Interestingly, this suggests that the efficiency boost provided by a scheduler in large-scale fuzzing tasks is comparable to the improvement offered by a fuzzer in a single program scenario. This is based on the fact that, for example, Google FuzzBench~\cite{metzman2021fuzzbench} reports an average increase of 20\%-30\% in coverage across multiple programs when using the best fuzzers compared to less effective ones~\cite{fuzzreport}.

\custombox{1}{ 
Advanced schedulers such as \tool lead to more effective fuzzing both in terms of overall code coverage across all programs and in terms of better coverage for majority of programs. }

\subsection{Scheduler vs Fuzzer}
\label{sect:eval:scheduler-vs-fuzzer}

In our second experiment, we aim to evaluate the significance of schedulers relative to fuzzers in the context of fuzzing at scale tasks. Specifically, we want to determine whether improving schedulers can yield a comparable impact to improving fuzzers. If this is the case, it would imply that schedulers deserve equal focus as fuzzers when developing more efficient fuzzing at scale frameworks.

We carry out three different fuzzing at scale tasks, independently on \benchmark and on \ossfuzz. In the first task, we use the baseline scheduler (round-robin) and the baseline fuzzer AFL. In the second, we switch the fuzzer from AFL to AFL++, but keep the same baseline scheduler. In the third, we keep AFL as a fuzzer, but replace the baseline scheduler with \tool. We then compare the outcomes of the three tasks on accumulative and voting metrics.

\begin{table}[h]
\caption{Comparison of importance of schedulers and fuzzers. The accumulative and vote columns represent the percentage increase in these two metrics between tasks in the first column and second columns. 
'RR' stands for round-robin (baseline) scheduler. 
}
\label{tbl:cmp:sch-fuzz}
\renewcommand{\arraystretch}{1.2}
\setlength\tabcolsep{4pt}
\begin{tabular}{|ll|ll|r|r|r|r|}
\multicolumn{4}{c}{ } & \multicolumn{2}{c}{\benchmark} & \multicolumn{2}{c}{\ossfuzz} \\
\hline
\multicolumn{2}{|c|}{Task 1}  & \multicolumn{2}{c|}{Task 2}   
& Accum &   Vote & Accum &   Vote \\
fuzz & sch & fuzz & sch & incr \%& incr \% & incr \%& incr \%\\
\hline
\hline
AFL   & RR & AFL++ & RR    & 14  & 11    & 11  & 10\\
AFL   & RR & AFL   & \tool & 41  & 12    & 10  & 3  \\
\hline
AFL++ & RR & AFL   & \tool & 23  & 1     &  0  & -11 \\
\hline
\end{tabular}%
\end{table}

The comparison results are presented in Table~\ref{tbl:cmp:sch-fuzz}. It is evident that using a superior fuzzer, such as replacing the standard AFL with the top-performer AFL++, while maintaining the baseline scheduler, leads to a 14\% increase in overall code coverage on \benchmark and 11\% on \ossfuzz. This means that AFL++ achieves, on average, 11-14\% more coverage than AFL. Furthermore, AFL++ offers better coverage for 10-11\% more programs than AFL. Thus, using a better fuzzer undoubtedly results in more efficient fuzzing at scale.
However, comparable and even superior results can be achieved simply by improving the scheduler, rather than the fuzzer. 
More precisely, by keeping the same fuzzer AFL, but utilizing \tool instead of the baseline scheduler, we observe 10-41\% increase in accumulative metric and  3-12\% increases in voting metric. 
In fact, the direct comparison between using an advanced fuzzer and baseline scheduler (AFL++ combined with round-robin), and baseline fuzzer and advanced scheduler (AFL combined with \tool) demonstrates that the latter provides higher overall coverage, whereas the former provides better coverage for more programs. 
We can thus conclude that, for effective fuzzing at scale, the scheduler is as crucial as the used fuzzer. Therefore, schedulers should receive a similar level of attention as fuzzers. However, they have been  overlooked thus far.

\custombox{2}{ 
Improving schedulers is as important as improving fuzzers. }

\subsection{Bugs in Ubuntu}
\label{sect:eval:bugs}
To underscore the importance of the fuzzing at scale framework, we further consider one practical task  -- identifying bugs in Ubuntu packages. 
For this purpose, we employ both AFL++ and Honggfuzz to fuzz \benchmark and discover bugs. We use \tool to schedule fuzzing at scale of the benchmark programs, independently with each fuzzer for 3 days on 30 CPUs, and collect all the bugs found and reported by the fuzzers. 
We intentionally choose not to separately report bugs found by AFL++ or Honggfuzz. This helps avoid unfair comparisons between the two fuzzers on bug-finding abilities which might have occurred due to our setup misconfigurations or the use of different sanitizers.
We merge the bugs identified by both fuzzers and deduplicate the resulting set. Specifically, we gather all inputs reported by the fuzzers that cause program crashes. Next, we verify that these inputs lead to crashes by executing the programs with the crashing inputs, ensuring that 
some signal is raised. 
Programs and inputs that pass this filter are then run with the dynamic binary analysis tool Valgrind~\cite{nethercote2007valgrind} to generate crash trace for each crash. We filter out duplicate traces and report all unique traces as individual bugs.
As a result, we find \tbugs bugs in \tprogs distinct programs among all the evaluated programs. Table~\ref{tbl:bugs} provides information on the classification of the discovered bugs.

\begin{table}[h]
\caption{Classification of the bugs found by fuzzing at scale \benchmark with \tool by using AFL++ and Honggfuzz for three days each on 30 CPUs. \code{OTHER} refers to all signals with signal number 32 and higher. %
}
\label{tbl:bugs}
\setlength{\tabcolsep}{0.1cm} 
\renewcommand{\arraystretch}{1.2}
\begin{tabular}{|l|l|r|}
\hline
Signal & Valgrind description & \# bugs \\
\hline
\hline
\code{SIGHUP}   &  \verb|<none>|  & 341 \\
\code{SIGHUP}   &  Access not within mapped region  & 4 \\
\hline
\code{SIGINT}   &  \verb|<none>| & 53 \\
\hline
\code{SIGQUIT}   &  \verb|<none>|   & 16 \\
\hline
\code{SIGILL}   &  \verb|<none>|   & 3 \\
\hline
\code{SIGTRAP}   &  \verb|<none>|   & 5 \\
\hline
\code{SIGABRT}   &  \verb|<none>|   & 1114 \\
\hline
\code{SIGBUS}   &  \verb|<none>|   & 5 \\
\code{SIGBUS}   &  Non-existent physical address   & 1 \\
\hline
\code{SIGFPE}   &  \verb|<none>|   & 20 \\
\code{SIGFPE}   &  Integer divide by zero   & 74 \\
\hline
\code{SIGKILL}   &  \verb|<none>|   & 3 \\
\hline
\code{SIGUSR1}   &  \verb|<none>|   & 1 \\
\hline
\code{SIGSEGV}   &  \verb|<none>|   & 321 \\
\code{SIGSEGV}   &  Access not within mapped region   & 1556 \\
\code{SIGSEGV}   &  Bad permissions for mapped region   & 878 \\
\code{SIGSEGV}   &  General Protection Fault   & 476 \\
\hline
\code{SIGUSR2}   &  \verb|<none>|   & 1 \\
\hline
\code{SIGXCPU}   &  \verb|<none>|   & 1 \\
\hline
\code{SIGXFSZ}   &  \verb|<none>|   & 4 \\
\hline
\code{SIGVTALRM}   &  \verb|<none>|   & 1 \\
\hline
\code{OTHER}   &  \verb|<none>|   & 30 \\
\hline

\end{tabular}%
\end{table}

\begin{figure}[h]
\centering
\includegraphics[scale=0.4]{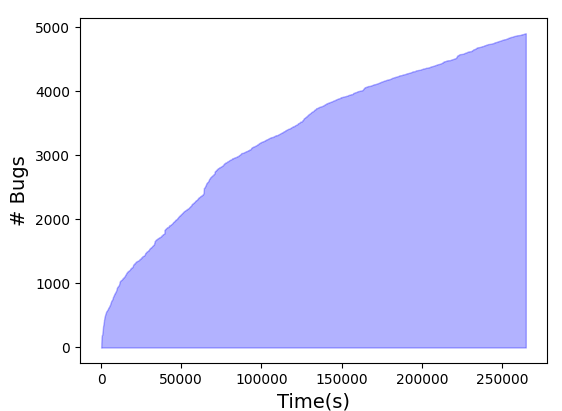}
\caption{The total number of found bugs in \benchmark during three-day fuzzing at scale campaign with AFL++ and Honggfuzz, each running on 30 CPUs.}
\label{fig:bugs:plot}
\end{figure}

The quantity of bugs identified during the three-day fuzzing session is fairly high. However, this number alone does not allow  to speculate whether extended fuzzing campaigns would yield a  larger number of bugs.
To address this, in Figure \ref{fig:bugs:plot} we  illustrate the cumulative count of discovered bugs as the fuzzing  progresses\footnote{We approximate the timing of each bug using the Linux timestamp associated with the update of the corresponding crash file.}. The plot strongly implies that the fuzzing process has not yet reached a plateau. Consequently, we can infer that extending the fuzzing period may result in a substantial increase in the number of found bugs.

\custombox{3}{ 
By utilizing fuzzing at scale for several days, we found  \tbugs bugs across \tprogs Ubuntu programs and longer fuzzing campaigns will likely lead to significantly more bugs.
}

Given the sheer volume of bugs identified, it is challenging to estimate their potential exploitability one-by-one. %
Instead, automated approaches are needed to identify vulnerabilities and report them. 
To seek assistance, we reached out to Ubuntu and Debian maintainers, but they indicated that no such approaches exist, as previously mentioned by a decade-old similar initiative~\cite{reportmayhem1,reportmayhem2} by project Mayhem~\cite{cha2012unleashing}.
We were advised to report the bugs directly to the developers and examine potential CVEs case-by-case. Currently, we are working on developing a strategy to automate these steps.

We did not run a similar fuzzing campaign for \ossfuzz programs because they have been fuzzed extensively on the Google infrastructure, so it is highly unlikely that a few days of fuzzing on our server would have produced new, previously unknown bugs.

\section{Related work}
\label{sect:related}

Platforms for fuzzing multiple programs distributed across multiple cores and machines have been developed by Google. 
OSS-Fuzz~\cite{serebryany2017oss} is a fuzzing platform for open-source programs. It is designed to find vulnerabilities in bulk, by continuously fuzzing programs on the Google infrastructure. It comes with around 1,000 open-source programs written in C, C++, Python, Go, JavaScript, Rust, Swift,  it supports AFL++, Honggfuzz, and libFuzzer fuzzers, and has automatic triage and reporting of bugs. As of February 2023, OSS-Fuzz has found around 28,000 bugs accross 850 projects~\cite{ossfuzzbugs}.
On the other hand,
ClusterFuzz~\cite{arya2019open} is the distributed fuzzing infrastructure behind OSS-Fuzz. It is open-source and allows for deployment within a personal environment. Note, OSS-Fuzz is a production instance of ClusterFuzz, however, it has some features that are not available in the open-source version. 
Neither OSS-Fuzz nor ClusterFuzz are 
wide 
fuzzing at scale frameworks. Their objective is more in line of distributing the fuzzing of smaller set of programs over vast resources.
In contrast, our goal is to optimize small resources for fuzzing of large number of programs.

In the context of fuzzers, the term 
\emph{scheduler} 
refers to the strategy that determines which testcase to choose for fuzzing, also known as seed scheduling. 
Various strategies exist for this purpose. Some employ a purely random selection~\cite{stephens2016driller}, while others utilize feedback from previous fuzzing sessions, such as seed size and execution time~\cite{zalewski2017american}, favoring less explored code regions~\cite{lemieux2018fairfuzz,bohme2016coverage}, applying supervised machine learning~\cite{chen2020meuzz}, or using other heuristics~\cite{dang2012rebucket,zhao2020probabilistic,she2022effective}. In contrast, our usage of the term \emph{scheduler} aligns more closely with its meaning in multitasking operating systems. In this sense, its primary goal is to manage the execution of different fuzzers on the available CPU resources.

Utilizing advanced techniques to guide selection in fuzzers is not a novel concept. For example, MOpt~\cite{lyu2019mopt} employs particle swarm optimization to steer the seed selection process. Meanwhile, EcoFuzz~~\cite{yue2020ecofuzz} leverages the multi-armed bandit (MAB) approach for seed selection. Sivo fuzzer~\cite{nikolic2021refined} also utilizes MABs, for guiding the selection of not just seeds, but other fuzzer components too. Our schedulers, particularly \tool, use MAB to assist in selecting the best fuzzer for execution on CPUs.

\section{Conclusion and future work}
\label{sect:conclusion}

In this study, we explored the problem of fuzzing at scale and demonstrated that non-trivial schedulers can impact its effectiveness. The impact of such schedulers is substantial, comparable to that of the fuzzers themselves. Our scheduler \tool is approximately 30\% more efficient than a basic scheduler. Further advancements in schedulers may lead to an even greater performance gap.
There are two prominent research directions that could potentially improve schedulers. The first direction is to concentrate solely on developing and using more advanced strategies for deciding which fuzzer to start based on the available past coverage information for each program. This could involve utilizing better multi-armed bandits or employing entirely new decision-making algorithms.
The second approach is to leverage the knowledge gained from fuzzing certain programs and applying it to fuzz other programs. For example, if a method can be devised to transfer the knowledge acquired from fuzzing one program (or a group of programs) to another similar program, the second programs may be fuzzed faster.
Advancing any of these two research directions will result in betters schedulers. This in turn will provide significant improvement in effectiveness of fuzzing at scale. 
We believe that the primary source of improvement will be derived from the progress in schedulers, as opposed to fuzzers, due to the relatively unexplored nature of the former and the well-established maturity of the latter.

\section*{Acknowledgements}
We thank Prateek Saxena and the anonymous reviewers for their valuable feedback.
This research was supported by the Ministry of Education
Singapore Tier 2 grant MOE-T2EP20220-0014.

\bibliographystyle{ACM-Reference-Format}
\bibliography{paper.bib}


\begin{thebibliography}{53}


\ifx \showCODEN    \undefined \def \showCODEN     #1{\unskip}     \fi
\ifx \showDOI      \undefined \def \showDOI       #1{#1}\fi
\ifx \showISBNx    \undefined \def \showISBNx     #1{\unskip}     \fi
\ifx \showISBNxiii \undefined \def \showISBNxiii  #1{\unskip}     \fi
\ifx \showISSN     \undefined \def \showISSN      #1{\unskip}     \fi
\ifx \showLCCN     \undefined \def \showLCCN      #1{\unskip}     \fi
\ifx \shownote     \undefined \def \shownote      #1{#1}          \fi
\ifx \showarticletitle \undefined \def \showarticletitle #1{#1}   \fi
\ifx \showURL      \undefined \def \showURL       {\relax}        \fi
\providecommand\bibfield[2]{#2}
\providecommand\bibinfo[2]{#2}
\providecommand\natexlab[1]{#1}
\providecommand\showeprint[2][]{arXiv:#2}

\bibitem[Arya et~al\mbox{.}(2019)]%
        {arya2019open}
\bibfield{author}{\bibinfo{person}{Abhishek Arya}, \bibinfo{person}{Oliver
  Chang}, \bibinfo{person}{Max Moroz}, \bibinfo{person}{Martin Barbella}, {and}
  \bibinfo{person}{Jonathan Metzman}.} \bibinfo{year}{2019}\natexlab{}.
\newblock \showarticletitle{Open sourcing Clusterfuzz}.
\newblock \bibinfo{journal}{\emph{Google, Inc. Feb}} (\bibinfo{year}{2019}).
\newblock


\bibitem[Aschermann et~al\mbox{.}(2019)]%
        {aschermann2019redqueen}
\bibfield{author}{\bibinfo{person}{Cornelius Aschermann},
  \bibinfo{person}{Sergej Schumilo}, \bibinfo{person}{Tim Blazytko},
  \bibinfo{person}{Robert Gawlik}, {and} \bibinfo{person}{Thorsten Holz}.}
  \bibinfo{year}{2019}\natexlab{}.
\newblock \showarticletitle{REDQUEEN: Fuzzing with Input-to-State
  Correspondence.}. In \bibinfo{booktitle}{\emph{NDSS}},
  Vol.~\bibinfo{volume}{19}. \bibinfo{pages}{1--15}.
\newblock


\bibitem[Blazytko et~al\mbox{.}(2019)]%
        {blazytko2019grimoire}
\bibfield{author}{\bibinfo{person}{Tim Blazytko}, \bibinfo{person}{Cornelius
  Aschermann}, \bibinfo{person}{Moritz Schl{\"o}gel}, \bibinfo{person}{Ali
  Abbasi}, \bibinfo{person}{Sergej Schumilo}, \bibinfo{person}{Simon
  W{\"o}rner}, {and} \bibinfo{person}{Thorsten Holz}.}
  \bibinfo{year}{2019}\natexlab{}.
\newblock \showarticletitle{GRIMOIRE: Synthesizing Structure while Fuzzing.}.
  In \bibinfo{booktitle}{\emph{USENIX Security Symposium}},
  Vol.~\bibinfo{volume}{19}.
\newblock


\bibitem[B{\"o}hme et~al\mbox{.}(2016)]%
        {bohme2016coverage}
\bibfield{author}{\bibinfo{person}{Marcel B{\"o}hme},
  \bibinfo{person}{Van-Thuan Pham}, {and} \bibinfo{person}{Abhik
  Roychoudhury}.} \bibinfo{year}{2016}\natexlab{}.
\newblock \showarticletitle{Coverage-based greybox fuzzing as markov chain}. In
  \bibinfo{booktitle}{\emph{Proceedings of the 2016 ACM SIGSAC Conference on
  Computer and Communications Security}}. \bibinfo{pages}{1032--1043}.
\newblock


\bibitem[Cha et~al\mbox{.}(2012)]%
        {cha2012unleashing}
\bibfield{author}{\bibinfo{person}{Sang~Kil Cha}, \bibinfo{person}{Thanassis
  Avgerinos}, \bibinfo{person}{Alexandre Rebert}, {and} \bibinfo{person}{David
  Brumley}.} \bibinfo{year}{2012}\natexlab{}.
\newblock \showarticletitle{Unleashing mayhem on binary code}. In
  \bibinfo{booktitle}{\emph{2012 IEEE Symposium on Security and Privacy}}.
  IEEE, \bibinfo{pages}{380--394}.
\newblock


\bibitem[Chen et~al\mbox{.}(2020b)]%
        {chen2020muzz}
\bibfield{author}{\bibinfo{person}{Hongxu Chen}, \bibinfo{person}{Shengjian
  Guo}, \bibinfo{person}{Yinxing Xue}, \bibinfo{person}{Yulei Sui},
  \bibinfo{person}{Cen Zhang}, \bibinfo{person}{Yuekang Li},
  \bibinfo{person}{Haijun Wang}, {and} \bibinfo{person}{Yang Liu}.}
  \bibinfo{year}{2020}\natexlab{b}.
\newblock \showarticletitle{MUZZ: Thread-aware grey-box fuzzing for effective
  bug hunting in multithreaded programs}.
\newblock \bibinfo{journal}{\emph{arXiv preprint arXiv:2007.15943}}
  (\bibinfo{year}{2020}).
\newblock


\bibitem[Chen and Chen(2018)]%
        {chen2018angora}
\bibfield{author}{\bibinfo{person}{Peng Chen} {and} \bibinfo{person}{Hao
  Chen}.} \bibinfo{year}{2018}\natexlab{}.
\newblock \showarticletitle{Angora: Efficient fuzzing by principled search}. In
  \bibinfo{booktitle}{\emph{2018 IEEE Symposium on Security and Privacy (SP)}}.
  IEEE, \bibinfo{pages}{711--725}.
\newblock


\bibitem[Chen et~al\mbox{.}(2019b)]%
        {chen2019matryoshka}
\bibfield{author}{\bibinfo{person}{Peng Chen}, \bibinfo{person}{Jianzhong Liu},
  {and} \bibinfo{person}{Hao Chen}.} \bibinfo{year}{2019}\natexlab{b}.
\newblock \showarticletitle{Matryoshka: fuzzing deeply nested branches}. In
  \bibinfo{booktitle}{\emph{Proceedings of the 2019 ACM SIGSAC Conference on
  Computer and Communications Security}}. \bibinfo{pages}{499--513}.
\newblock


\bibitem[Chen et~al\mbox{.}(2020a)]%
        {chen2020meuzz}
\bibfield{author}{\bibinfo{person}{Yaohui Chen}, \bibinfo{person}{Mansour
  Ahmadi}, \bibinfo{person}{Reza~Mirzazade Farkhani}, \bibinfo{person}{Boyu
  Wang}, {and} \bibinfo{person}{Long Lu}.} \bibinfo{year}{2020}\natexlab{a}.
\newblock \showarticletitle{MEUZZ: Smart Seed Scheduling for Hybrid Fuzzing.}.
  In \bibinfo{booktitle}{\emph{RAID}}. \bibinfo{pages}{77--92}.
\newblock


\bibitem[Chen et~al\mbox{.}(2019a)]%
        {chen2019enfuzz}
\bibfield{author}{\bibinfo{person}{Yuanliang Chen}, \bibinfo{person}{Yu Jiang},
  \bibinfo{person}{Fuchen Ma}, \bibinfo{person}{Jie Liang},
  \bibinfo{person}{Mingzhe Wang}, \bibinfo{person}{Chijin Zhou},
  \bibinfo{person}{Xun Jiao}, {and} \bibinfo{person}{Zhuo Su}.}
  \bibinfo{year}{2019}\natexlab{a}.
\newblock \showarticletitle{EnFuzz: Ensemble Fuzzing with Seed Synchronization
  among Diverse Fuzzers.}. In \bibinfo{booktitle}{\emph{USENIX Security
  Symposium}}. \bibinfo{pages}{1967--1983}.
\newblock


\bibitem[Cho et~al\mbox{.}(2019)]%
        {cho2019intriguer}
\bibfield{author}{\bibinfo{person}{Mingi Cho}, \bibinfo{person}{Seoyoung Kim},
  {and} \bibinfo{person}{Taekyoung Kwon}.} \bibinfo{year}{2019}\natexlab{}.
\newblock \showarticletitle{Intriguer: Field-level constraint solving for
  hybrid fuzzing}. In \bibinfo{booktitle}{\emph{Proceedings of the 2019 ACM
  SIGSAC Conference on Computer and Communications Security}}.
  \bibinfo{pages}{515--530}.
\newblock


\bibitem[Dang et~al\mbox{.}(2012)]%
        {dang2012rebucket}
\bibfield{author}{\bibinfo{person}{Yingnong Dang}, \bibinfo{person}{Rongxin
  Wu}, \bibinfo{person}{Hongyu Zhang}, \bibinfo{person}{Dongmei Zhang}, {and}
  \bibinfo{person}{Peter Nobel}.} \bibinfo{year}{2012}\natexlab{}.
\newblock \showarticletitle{Rebucket: A method for clustering duplicate crash
  reports based on call stack similarity}. In \bibinfo{booktitle}{\emph{2012
  34th International Conference on Software Engineering (ICSE)}}. IEEE,
  \bibinfo{pages}{1084--1093}.
\newblock


\bibitem[Debian(2013)]%
        {reportmayhem1}
\bibfield{author}{\bibinfo{person}{Debian}.} \bibinfo{year}{2013}\natexlab{}.
\newblock \bibinfo{title}{Reporting 1.2K crashes}.
\newblock
  \bibinfo{howpublished}{\url{https://lists.debian.org/debian-devel/2013/06/msg00720.html}}.
\newblock
\newblock
\shownote{Accessed: May 17, 2023}.


\bibitem[Fioraldi et~al\mbox{.}(2020)]%
        {fioraldi2020afl}
\bibfield{author}{\bibinfo{person}{Andrea Fioraldi}, \bibinfo{person}{Dominik
  Maier}, \bibinfo{person}{Heiko Ei{\ss}feldt}, {and} \bibinfo{person}{Marc
  Heuse}.} \bibinfo{year}{2020}\natexlab{}.
\newblock \showarticletitle{{AFL++} combining incremental steps of fuzzing
  research}. In \bibinfo{booktitle}{\emph{Proceedings of the 14th USENIX
  Conference on Offensive Technologies}}. \bibinfo{pages}{10--10}.
\newblock


\bibitem[Gan et~al\mbox{.}(2020)]%
        {gan2020greyone}
\bibfield{author}{\bibinfo{person}{Shuitao Gan}, \bibinfo{person}{Chao Zhang},
  \bibinfo{person}{Peng Chen}, \bibinfo{person}{Bodong Zhao},
  \bibinfo{person}{Xiaojun Qin}, \bibinfo{person}{Dong Wu}, {and}
  \bibinfo{person}{Zuoning Chen}.} \bibinfo{year}{2020}\natexlab{}.
\newblock \showarticletitle{GREYONE: Data Flow Sensitive Fuzzing.}. In
  \bibinfo{booktitle}{\emph{USENIX Security Symposium}}.
  \bibinfo{pages}{2577--2594}.
\newblock


\bibitem[Gan et~al\mbox{.}(2018)]%
        {gan2018collafl}
\bibfield{author}{\bibinfo{person}{Shuitao Gan}, \bibinfo{person}{Chao Zhang},
  \bibinfo{person}{Xiaojun Qin}, \bibinfo{person}{Xuwen Tu},
  \bibinfo{person}{Kang Li}, \bibinfo{person}{Zhongyu Pei}, {and}
  \bibinfo{person}{Zuoning Chen}.} \bibinfo{year}{2018}\natexlab{}.
\newblock \showarticletitle{Collafl: Path sensitive fuzzing}. In
  \bibinfo{booktitle}{\emph{2018 IEEE Symposium on Security and Privacy (SP)}}.
  IEEE, \bibinfo{pages}{679--696}.
\newblock


\bibitem[Geretto et~al\mbox{.}(2022)]%
        {geretto2022snappy}
\bibfield{author}{\bibinfo{person}{Elia Geretto}, \bibinfo{person}{Cristiano
  Giuffrida}, \bibinfo{person}{Herbert Bos}, {and} \bibinfo{person}{Erik Van
  Der~Kouwe}.} \bibinfo{year}{2022}\natexlab{}.
\newblock \showarticletitle{Snappy: Efficient Fuzzing with Adaptive and Mutable
  Snapshots}. In \bibinfo{booktitle}{\emph{Proceedings of the 38th Annual
  Computer Security Applications Conference}}. \bibinfo{pages}{375--387}.
\newblock


\bibitem[Google(2022)]%
        {fuzzreport}
\bibfield{author}{\bibinfo{person}{Google}.} \bibinfo{year}{2022}\natexlab{}.
\newblock \bibinfo{title}{FuzzBench: 2022-09-29 report}.
\newblock
  \bibinfo{howpublished}{\url{https://www.fuzzbench.com/reports/2022-09-29/index.html}}.
\newblock
\newblock
\shownote{Accessed: May 17, 2023}.


\bibitem[Google(2023)]%
        {ossfuzzbugs}
\bibfield{author}{\bibinfo{person}{Google}.} \bibinfo{year}{2023}\natexlab{}.
\newblock \bibinfo{title}{{OSS}-Fuzz}.
\newblock \bibinfo{howpublished}{\url{https://google.github.io/oss-fuzz/}}.
\newblock
\newblock
\shownote{Accessed: May 17, 2023}.


\bibitem[Koike et~al\mbox{.}(2022)]%
        {koike2022slopt}
\bibfield{author}{\bibinfo{person}{Yuki Koike}, \bibinfo{person}{Hiroyuki
  Katsura}, \bibinfo{person}{Hiromu Yakura}, {and} \bibinfo{person}{Yuma
  Kurogome}.} \bibinfo{year}{2022}\natexlab{}.
\newblock \showarticletitle{SLOPT: Bandit Optimization Framework for
  Mutation-Based Fuzzing}. In \bibinfo{booktitle}{\emph{Proceedings of the 38th
  Annual Computer Security Applications Conference}}.
  \bibinfo{pages}{519--533}.
\newblock


\bibitem[Lee et~al\mbox{.}(2021)]%
        {lee2021constraint}
\bibfield{author}{\bibinfo{person}{Gwangmu Lee}, \bibinfo{person}{Woochul
  Shim}, {and} \bibinfo{person}{Byoungyoung Lee}.}
  \bibinfo{year}{2021}\natexlab{}.
\newblock \showarticletitle{Constraint-guided Directed Greybox Fuzzing.}. In
  \bibinfo{booktitle}{\emph{USENIX Security Symposium}}.
  \bibinfo{pages}{3559--3576}.
\newblock


\bibitem[Lemieux and Sen(2018)]%
        {lemieux2018fairfuzz}
\bibfield{author}{\bibinfo{person}{Caroline Lemieux} {and}
  \bibinfo{person}{Koushik Sen}.} \bibinfo{year}{2018}\natexlab{}.
\newblock \showarticletitle{Fairfuzz: A targeted mutation strategy for
  increasing greybox fuzz testing coverage}. In
  \bibinfo{booktitle}{\emph{Proceedings of the 33rd ACM/IEEE International
  Conference on Automated Software Engineering}}. \bibinfo{pages}{475--485}.
\newblock


\bibitem[Liang et~al\mbox{.}(2019)]%
        {liang2019deepfuzzer}
\bibfield{author}{\bibinfo{person}{Jie Liang}, \bibinfo{person}{Yu Jiang},
  \bibinfo{person}{Mingzhe Wang}, \bibinfo{person}{Xun Jiao},
  \bibinfo{person}{Yuanliang Chen}, \bibinfo{person}{Houbing Song}, {and}
  \bibinfo{person}{Kim-Kwang~Raymond Choo}.} \bibinfo{year}{2019}\natexlab{}.
\newblock \showarticletitle{Deepfuzzer: Accelerated deep greybox fuzzing}.
\newblock \bibinfo{journal}{\emph{IEEE Transactions on Dependable and Secure
  Computing}} \bibinfo{volume}{18}, \bibinfo{number}{6} (\bibinfo{year}{2019}),
  \bibinfo{pages}{2675--2688}.
\newblock


\bibitem[Liang et~al\mbox{.}(2022)]%
        {liang2022pata}
\bibfield{author}{\bibinfo{person}{Jie Liang}, \bibinfo{person}{Mingzhe Wang},
  \bibinfo{person}{Chijin Zhou}, \bibinfo{person}{Zhiyong Wu},
  \bibinfo{person}{Yu Jiang}, \bibinfo{person}{Jianzhong Liu},
  \bibinfo{person}{Zhe Liu}, {and} \bibinfo{person}{Jiaguang Sun}.}
  \bibinfo{year}{2022}\natexlab{}.
\newblock \showarticletitle{PATA: Fuzzing with path aware taint analysis}. In
  \bibinfo{booktitle}{\emph{2022 IEEE Symposium on Security and Privacy (SP)}}.
  IEEE, \bibinfo{pages}{1--17}.
\newblock


\bibitem[Liu and Layland(1973)]%
        {liu1973scheduling}
\bibfield{author}{\bibinfo{person}{Chung~Laung Liu} {and}
  \bibinfo{person}{James~W Layland}.} \bibinfo{year}{1973}\natexlab{}.
\newblock \showarticletitle{Scheduling algorithms for multiprogramming in a
  hard-real-time environment}.
\newblock \bibinfo{journal}{\emph{Journal of the ACM (JACM)}}
  \bibinfo{volume}{20}, \bibinfo{number}{1} (\bibinfo{year}{1973}),
  \bibinfo{pages}{46--61}.
\newblock


\bibitem[lwn.net(2013)]%
        {reportmayhem2}
\bibfield{author}{\bibinfo{person}{lwn.net}.} \bibinfo{year}{2013}\natexlab{}.
\newblock \bibinfo{title}{Mayhem finds 1200 bugs}.
\newblock \bibinfo{howpublished}{\url{https://lwn.net/Articles/557055/}}.
\newblock
\newblock
\shownote{Accessed: May 17, 2023}.


\bibitem[Lyu et~al\mbox{.}(2019)]%
        {lyu2019mopt}
\bibfield{author}{\bibinfo{person}{Chenyang Lyu}, \bibinfo{person}{Shouling
  Ji}, \bibinfo{person}{Chao Zhang}, \bibinfo{person}{Yuwei Li},
  \bibinfo{person}{Wei-Han Lee}, \bibinfo{person}{Yu Song}, {and}
  \bibinfo{person}{Raheem Beyah}.} \bibinfo{year}{2019}\natexlab{}.
\newblock \showarticletitle{MOPT: Optimized Mutation Scheduling for Fuzzers.}.
  In \bibinfo{booktitle}{\emph{USENIX Security Symposium}}.
  \bibinfo{pages}{1949--1966}.
\newblock


\bibitem[Lyu et~al\mbox{.}(2022)]%
        {lyu2022ems}
\bibfield{author}{\bibinfo{person}{Chenyang Lyu}, \bibinfo{person}{Shouling
  Ji}, \bibinfo{person}{Xuhong Zhang}, \bibinfo{person}{Hong Liang},
  \bibinfo{person}{Binbin Zhao}, \bibinfo{person}{Kangjie Lu}, {and}
  \bibinfo{person}{Raheem Beyah}.} \bibinfo{year}{2022}\natexlab{}.
\newblock \showarticletitle{Ems: History-driven mutation for coverage-based
  fuzzing}. In \bibinfo{booktitle}{\emph{29rd Annual Network and Distributed
  System Security Symposium, NDSS}}. \bibinfo{pages}{24--28}.
\newblock


\bibitem[Metzman et~al\mbox{.}(2021)]%
        {metzman2021fuzzbench}
\bibfield{author}{\bibinfo{person}{Jonathan Metzman},
  \bibinfo{person}{L{\'a}szl{\'o} Szekeres}, \bibinfo{person}{Laurent Simon},
  \bibinfo{person}{Read Sprabery}, {and} \bibinfo{person}{Abhishek Arya}.}
  \bibinfo{year}{2021}\natexlab{}.
\newblock \showarticletitle{Fuzzbench: an open fuzzer benchmarking platform and
  service}. In \bibinfo{booktitle}{\emph{Proceedings of the 29th ACM joint
  meeting on European software engineering conference and symposium on the
  foundations of software engineering}}. \bibinfo{pages}{1393--1403}.
\newblock


\bibitem[Nagy et~al\mbox{.}(2021a)]%
        {nagy2021breaking}
\bibfield{author}{\bibinfo{person}{Stefan Nagy}, \bibinfo{person}{Anh
  Nguyen-Tuong}, \bibinfo{person}{Jason~D Hiser}, \bibinfo{person}{Jack~W
  Davidson}, {and} \bibinfo{person}{Matthew Hicks}.}
  \bibinfo{year}{2021}\natexlab{a}.
\newblock \showarticletitle{Breaking through binaries: Compiler-quality
  instrumentation for better binary-only fuzzing}. In
  \bibinfo{booktitle}{\emph{30th USENIX Security Symposium}}.
\newblock


\bibitem[Nagy et~al\mbox{.}(2021b)]%
        {nagy2021same}
\bibfield{author}{\bibinfo{person}{Stefan Nagy}, \bibinfo{person}{Anh
  Nguyen-Tuong}, \bibinfo{person}{Jason~D Hiser}, \bibinfo{person}{Jack~W
  Davidson}, {and} \bibinfo{person}{Matthew Hicks}.}
  \bibinfo{year}{2021}\natexlab{b}.
\newblock \showarticletitle{Same Coverage, Less Bloat: Accelerating Binary-only
  Fuzzing with Coverage-preserving Coverage-guided Tracing}. In
  \bibinfo{booktitle}{\emph{Proceedings of the 2021 ACM SIGSAC Conference on
  Computer and Communications Security}}. \bibinfo{pages}{351--365}.
\newblock


\bibitem[Nethercote and Seward(2007)]%
        {nethercote2007valgrind}
\bibfield{author}{\bibinfo{person}{Nicholas Nethercote} {and}
  \bibinfo{person}{Julian Seward}.} \bibinfo{year}{2007}\natexlab{}.
\newblock \showarticletitle{Valgrind: a framework for heavyweight dynamic
  binary instrumentation}.
\newblock \bibinfo{journal}{\emph{ACM Sigplan notices}} \bibinfo{volume}{42},
  \bibinfo{number}{6} (\bibinfo{year}{2007}), \bibinfo{pages}{89--100}.
\newblock


\bibitem[Nguyen et~al\mbox{.}(2020)]%
        {nguyen2020binary}
\bibfield{author}{\bibinfo{person}{Manh-Dung Nguyen},
  \bibinfo{person}{S{\'e}bastien Bardin}, \bibinfo{person}{Richard Bonichon},
  \bibinfo{person}{Roland Groz}, {and} \bibinfo{person}{Matthieu Lemerre}.}
  \bibinfo{year}{2020}\natexlab{}.
\newblock \showarticletitle{Binary-level Directed Fuzzing for Use-After-Free
  Vulnerabilities.}. In \bibinfo{booktitle}{\emph{RAID}}.
  \bibinfo{pages}{47--62}.
\newblock


\bibitem[Nikoli{\'c} et~al\mbox{.}(2021)]%
        {nikolic2021refined}
\bibfield{author}{\bibinfo{person}{Ivica Nikoli{\'c}}, \bibinfo{person}{Radu
  Mantu}, \bibinfo{person}{Shiqi Shen}, {and} \bibinfo{person}{Prateek
  Saxena}.} \bibinfo{year}{2021}\natexlab{}.
\newblock \showarticletitle{Refined grey-box fuzzing with Sivo}. In
  \bibinfo{booktitle}{\emph{Detection of Intrusions and Malware, and
  Vulnerability Assessment: 18th International Conference, DIMVA 2021, Virtual
  Event, July 14--16, 2021, Proceedings 18}}. Springer,
  \bibinfo{pages}{106--129}.
\newblock


\bibitem[{\"O}sterlund et~al\mbox{.}(2020)]%
        {osterlund2020parmesan}
\bibfield{author}{\bibinfo{person}{Sebastian {\"O}sterlund},
  \bibinfo{person}{Kaveh Razavi}, \bibinfo{person}{Herbert Bos}, {and}
  \bibinfo{person}{Cristiano Giuffrida}.} \bibinfo{year}{2020}\natexlab{}.
\newblock \showarticletitle{Parmesan: Sanitizer-guided greybox fuzzing}. In
  \bibinfo{booktitle}{\emph{Proceedings of the 29th USENIX Conference on
  Security Symposium}}. \bibinfo{pages}{2289--2306}.
\newblock


\bibitem[Peterson and Silberschatz(1985)]%
        {peterson1985operating}
\bibfield{author}{\bibinfo{person}{James~L Peterson} {and}
  \bibinfo{person}{Abraham Silberschatz}.} \bibinfo{year}{1985}\natexlab{}.
\newblock \bibinfo{booktitle}{\emph{Operating system concepts}}.
\newblock \bibinfo{publisher}{Addison-Wesley Longman Publishing Co., Inc.}
\newblock


\bibitem[Rawat et~al\mbox{.}(2017)]%
        {rawat2017vuzzer}
\bibfield{author}{\bibinfo{person}{Sanjay Rawat}, \bibinfo{person}{Vivek Jain},
  \bibinfo{person}{Ashish Kumar}, \bibinfo{person}{Lucian Cojocar},
  \bibinfo{person}{Cristiano Giuffrida}, {and} \bibinfo{person}{Herbert Bos}.}
  \bibinfo{year}{2017}\natexlab{}.
\newblock \showarticletitle{Vuzzer: Application-aware evolutionary fuzzing.}.
  In \bibinfo{booktitle}{\emph{NDSS}}, Vol.~\bibinfo{volume}{17}.
  \bibinfo{pages}{1--14}.
\newblock


\bibitem[Serebryany(2017)]%
        {serebryany2017oss}
\bibfield{author}{\bibinfo{person}{Kostya Serebryany}.}
  \bibinfo{year}{2017}\natexlab{}.
\newblock \showarticletitle{{OSS-Fuzz}-{G}oogle’s continuous fuzzing service
  for open source software}. In \bibinfo{booktitle}{\emph{USENIX Security
  symposium}}. USENIX Association.
\newblock


\bibitem[Shah et~al\mbox{.}(2022)]%
        {shah2022mc2}
\bibfield{author}{\bibinfo{person}{Abhishek Shah}, \bibinfo{person}{Dongdong
  She}, \bibinfo{person}{Samanway Sadhu}, \bibinfo{person}{Krish Singal},
  \bibinfo{person}{Peter Coffman}, {and} \bibinfo{person}{Suman Jana}.}
  \bibinfo{year}{2022}\natexlab{}.
\newblock \showarticletitle{MC2: Rigorous and Efficient Directed Greybox
  Fuzzing}. In \bibinfo{booktitle}{\emph{Proceedings of the 2022 ACM SIGSAC
  Conference on Computer and Communications Security}}.
  \bibinfo{pages}{2595--2609}.
\newblock


\bibitem[She et~al\mbox{.}(2022)]%
        {she2022effective}
\bibfield{author}{\bibinfo{person}{Dongdong She}, \bibinfo{person}{Abhishek
  Shah}, {and} \bibinfo{person}{Suman Jana}.} \bibinfo{year}{2022}\natexlab{}.
\newblock \showarticletitle{Effective seed scheduling for fuzzing with graph
  centrality analysis}. In \bibinfo{booktitle}{\emph{2022 IEEE Symposium on
  Security and Privacy (SP)}}. IEEE, \bibinfo{pages}{2194--2211}.
\newblock


\bibitem[Slivkins et~al\mbox{.}(2019)]%
        {slivkins2019introduction}
\bibfield{author}{\bibinfo{person}{Aleksandrs Slivkins} {et~al\mbox{.}}}
  \bibinfo{year}{2019}\natexlab{}.
\newblock \showarticletitle{Introduction to multi-armed bandits}.
\newblock \bibinfo{journal}{\emph{Foundations and Trends{\textregistered} in
  Machine Learning}} \bibinfo{volume}{12}, \bibinfo{number}{1-2}
  (\bibinfo{year}{2019}), \bibinfo{pages}{1--286}.
\newblock


\bibitem[Stephens et~al\mbox{.}(2016)]%
        {stephens2016driller}
\bibfield{author}{\bibinfo{person}{Nick Stephens}, \bibinfo{person}{John
  Grosen}, \bibinfo{person}{Christopher Salls}, \bibinfo{person}{Andrew
  Dutcher}, \bibinfo{person}{Ruoyu Wang}, \bibinfo{person}{Jacopo Corbetta},
  \bibinfo{person}{Yan Shoshitaishvili}, \bibinfo{person}{Christopher Kruegel},
  {and} \bibinfo{person}{Giovanni Vigna}.} \bibinfo{year}{2016}\natexlab{}.
\newblock \showarticletitle{Driller: Augmenting fuzzing through selective
  symbolic execution.}. In \bibinfo{booktitle}{\emph{NDSS}},
  Vol.~\bibinfo{volume}{16}. \bibinfo{pages}{1--16}.
\newblock


\bibitem[Sutton and Barto(2018)]%
        {sutton2018reinforcement}
\bibfield{author}{\bibinfo{person}{Richard~S Sutton} {and}
  \bibinfo{person}{Andrew~G Barto}.} \bibinfo{year}{2018}\natexlab{}.
\newblock \bibinfo{booktitle}{\emph{Reinforcement learning: An introduction}}.
\newblock \bibinfo{publisher}{MIT press}.
\newblock


\bibitem[Swiecki et~al\mbox{.}(2021)]%
        {swiecki2021honggfuzz}
\bibfield{author}{\bibinfo{person}{Robert Swiecki} {et~al\mbox{.}}}
  \bibinfo{year}{2021}\natexlab{}.
\newblock \bibinfo{title}{Honggfuzz-Security oriented software fuzzer}.
\newblock
\newblock


\bibitem[Wang et~al\mbox{.}({[n.\,d.]})]%
        {wangcarpetfuzz}
\bibfield{author}{\bibinfo{person}{Dawei Wang}, \bibinfo{person}{Ying Li},
  \bibinfo{person}{Zhiyu Zhang}, {and} \bibinfo{person}{Kai Chen}.}
  \bibinfo{year}{[n.\,d.]}\natexlab{}.
\newblock \showarticletitle{CarpetFuzz: Automatic Program Option Constraint
  Extraction from Documentation for Fuzzing}.
\newblock  (\bibinfo{year}{[n.\,d.]}).
\newblock


\bibitem[Yu et~al\mbox{.}(2022)]%
        {yu2022htfuzz}
\bibfield{author}{\bibinfo{person}{Yuanping Yu}, \bibinfo{person}{Xiangkun
  Jia}, \bibinfo{person}{Yuwei Liu}, \bibinfo{person}{Yanhao Wang},
  \bibinfo{person}{Qian Sang}, \bibinfo{person}{Chao Zhang}, {and}
  \bibinfo{person}{Purui Su}.} \bibinfo{year}{2022}\natexlab{}.
\newblock \showarticletitle{HTFuzz: Heap Operation Sequence Sensitive Fuzzing}.
  In \bibinfo{booktitle}{\emph{Proceedings of the 37th IEEE/ACM International
  Conference on Automated Software Engineering}}. \bibinfo{pages}{1--13}.
\newblock


\bibitem[Yue et~al\mbox{.}(2020)]%
        {yue2020ecofuzz}
\bibfield{author}{\bibinfo{person}{Tai Yue}, \bibinfo{person}{Pengfei Wang},
  \bibinfo{person}{Yong Tang}, \bibinfo{person}{Enze Wang}, \bibinfo{person}{Bo
  Yu}, \bibinfo{person}{Kai Lu}, {and} \bibinfo{person}{Xu Zhou}.}
  \bibinfo{year}{2020}\natexlab{}.
\newblock \showarticletitle{Ecofuzz: Adaptive energy-saving greybox fuzzing as
  a variant of the adversarial multi-armed bandit}. In
  \bibinfo{booktitle}{\emph{Proceedings of the 29th USENIX Conference on
  Security Symposium}}. \bibinfo{pages}{2307--2324}.
\newblock


\bibitem[Yun et~al\mbox{.}(2018)]%
        {yun2018qsym}
\bibfield{author}{\bibinfo{person}{Insu Yun}, \bibinfo{person}{Sangho Lee},
  \bibinfo{person}{Meng Xu}, \bibinfo{person}{Yeongjin Jang}, {and}
  \bibinfo{person}{Taesoo Kim}.} \bibinfo{year}{2018}\natexlab{}.
\newblock \showarticletitle{$\{$QSYM$\}$: A practical concolic execution engine
  tailored for hybrid fuzzing}. In \bibinfo{booktitle}{\emph{27th
  $\{$USENIX$\}$ Security Symposium ($\{$USENIX$\}$ Security 18)}}.
  \bibinfo{pages}{745--761}.
\newblock


\bibitem[Zalewski(2017)]%
        {zalewski2017american}
\bibfield{author}{\bibinfo{person}{Michal Zalewski}.}
  \bibinfo{year}{2017}\natexlab{}.
\newblock \bibinfo{title}{American fuzzy lop (AFL) fuzzer}.
\newblock
\newblock


\bibitem[Zhao et~al\mbox{.}(2020)]%
        {zhao2020probabilistic}
\bibfield{author}{\bibinfo{person}{Lei Zhao}, \bibinfo{person}{Pengcheng Cao},
  \bibinfo{person}{Yue Duan}, \bibinfo{person}{Heng Yin}, {and}
  \bibinfo{person}{Jifeng Xuan}.} \bibinfo{year}{2020}\natexlab{}.
\newblock \showarticletitle{Probabilistic path prioritization for hybrid
  fuzzing}.
\newblock \bibinfo{journal}{\emph{IEEE Transactions on Dependable and Secure
  Computing}} \bibinfo{volume}{19}, \bibinfo{number}{3} (\bibinfo{year}{2020}),
  \bibinfo{pages}{1955--1973}.
\newblock


\bibitem[Zhou et~al\mbox{.}(2022)]%
        {zhou2022ferry}
\bibfield{author}{\bibinfo{person}{Shunfan Zhou}, \bibinfo{person}{Zhemin
  Yang}, \bibinfo{person}{Dan Qiao}, \bibinfo{person}{Peng Liu},
  \bibinfo{person}{Min Yang}, \bibinfo{person}{Zhe Wang}, {and}
  \bibinfo{person}{Chenggang Wu}.} \bibinfo{year}{2022}\natexlab{}.
\newblock \showarticletitle{Ferry:$\{$State-Aware$\}$ Symbolic Execution for
  Exploring $\{$State-Dependent$\}$ Program Paths}. In
  \bibinfo{booktitle}{\emph{31st USENIX Security Symposium (USENIX Security
  22)}}. \bibinfo{pages}{4365--4382}.
\newblock


\bibitem[Zhu and B{\"o}hme(2021)]%
        {zhu2021regression}
\bibfield{author}{\bibinfo{person}{Xiaogang Zhu} {and} \bibinfo{person}{Marcel
  B{\"o}hme}.} \bibinfo{year}{2021}\natexlab{}.
\newblock \showarticletitle{Regression greybox fuzzing}. In
  \bibinfo{booktitle}{\emph{Proceedings of the 2021 ACM SIGSAC Conference on
  Computer and Communications Security}}. \bibinfo{pages}{2169--2182}.
\newblock


\bibitem[Zong et~al\mbox{.}(2020)]%
        {zong2020fuzzguard}
\bibfield{author}{\bibinfo{person}{Peiyuan Zong}, \bibinfo{person}{Tao Lv},
  \bibinfo{person}{Dawei Wang}, \bibinfo{person}{Zizhuang Deng},
  \bibinfo{person}{Ruigang Liang}, {and} \bibinfo{person}{Kai Chen}.}
  \bibinfo{year}{2020}\natexlab{}.
\newblock \showarticletitle{Fuzzguard: Filtering out unreachable inputs in
  directed grey-box fuzzing through deep learning}. In
  \bibinfo{booktitle}{\emph{Proceedings of the 29th USENIX Conference on
  Security Symposium}}. \bibinfo{pages}{2255--2269}.
\newblock


\end{thebibliography}

\end{document}